\documentclass[%
reprint,aps,amsmath,amssymb,prb,superscriptaddress,longbibliography%
]{revtex4-2}
\usepackage{dcolumn} 
\usepackage{graphicx} 
\usepackage{graphicx,subcaption}
\usepackage[version=3]{mhchem}
\usepackage{color,soul}
\usepackage{multirow}
\usepackage{xcolor}
\usepackage{bm}
\usepackage{comment}
\usepackage{float}
\usepackage{amsmath}
\usepackage{physics}
\usepackage{accents} 
\usepackage{xcolor}
\usepackage{bm}
\usepackage{comment}
\usepackage{blkarray}
\usepackage{hyperref}

\newcommand{\joe}[1]{\textcolor{black}{#1}}
\newcommand{\nadine}[1]{\textcolor{black}{#1}}

\newcommand{\up}{{{p}}}

\newcommand{\bR}{{{\bm R}}}

\newcommand{\bL}{{{\bm L}}}
\newcommand{\bP}{{{\bm P}}}
\newcommand{\hbm}[1]{\hat{\bm{#1}}}
\newcommand{\uvR}{{\bm{R}}}
\newcommand{\uvX}{{\bm{X}}}
\newcommand{\uvr}{{\bm{r}}}

\newcommand{\uvP}{{\bm{P}}}
\newcommand{\bd}{{\bm{d}}}
\newcommand{\uvp}{{\bm{p}}}
\newcommand{\uvPR}{{\bm{P}}_R}
\newcommand{\uvPr}{{\bm{P}}_r}
\newcommand{\uvPC}{{\bm{P}}_{\MCM}}

\newcommand{\bvR}{{\mathbf{R}}}
\newcommand{\bvP}{{\mathbf{P}}}
\newcommand{\hG}{\hat{\Gamma}}
\newcommand{\hbp}{\hat{\bm{p}}}
\newcommand{\hbl}{\hat{\bm{l}}}
\newcommand{\hbP}{\hat{\bm{P}}}
\newcommand{\hbr}{\hat{\bm{r}}}
\newcommand{\hbG}{\hat{\bm{\Gamma}}}
\newcommand{\hV}{\hat{V}}
\newcommand{\hH}{\hat{H}}

\newcommand{\MCM}{\textrm{MCM}}
\newcommand{\NCM}{\textrm{NCM}}
\newcommand{\TOT}{\textrm{TOT}}

\newcommand{\T}{\textrm{T}}
\newcommand{\lf}{\textrm{lf}} 
\newcommand{\bff}{\textrm{bf}}
\newcommand{\Int}{\textrm{int}}
\newcommand{\el}{\textrm{el}}
\newcommand{\ad}{\textrm{ad}}
\newcommand{\PS}{\textrm{PS}}
\newcommand{\BO}{\textrm{BO}}
\newcommand{\vib}{\textrm{vib}}
\newcommand{\DBOC}{\textrm{DBOC}}

\begin{document}
\title{A phase space electronic structure theory beyond Born-Oppenheimer: Geometric forces and Moody-Shapere-Wilczek  revisited}

\author{Mansi Bhati}
\affiliation{Department of Chemistry, Princeton University, Princeton, NJ USA}
\author{D. Vale Cofer-Shabica}
\affiliation{Department of Chemistry, Princeton University, Princeton, NJ USA}
\author{Jonathan I. Rawlinson}
\affiliation{Department of Mathematics, Nottingham Trent University, Nottingham, United Kingdom}
\author{Robert G. Littlejohn}
\affiliation{Department of Physics, University of California Berkeley, Berkeley, CA USA}
\author{Joseph Subotnik}
\email{subotnik@princeton.edu}
\affiliation{Department of Chemistry, Princeton University, Princeton, NJ USA}
\author{Nadine C. Bradbury}
\email{nadinebradbury@princeton.edu}
\affiliation{Department of Chemistry, Princeton University, Princeton, NJ USA}

\begin{abstract}
    We revisit the three-body problem in quantum mechanics in two and three dimensions, generating both exact eigenvalues and eigenvectors of the Hamiltonian and a series of  approximate solutions as calculated with a variety of different schemes to separate heavy (``nuclear") and light (``electronic") particles. We show that, with minimal extra cost, one can go beyond the Born-Oppenheimer approximation by performing electronic structure calculations parameterized by both the nuclear position (${\bm X})$ and the nuclear momentum ($\bm{P}$), a so-called phase space theory of electronic structure. In particular, we demonstrate that such phase space electronic structure calculations correctly incorporate the non-inertial Coriolis and centrifugal forces felt by electrons in a moving nuclear frame, thus leading to far more accurate eigenenergies and electronic angular momenta than has been possible before. We also demonstrate that our approach naturally incorporates and generalizes the Moody-Shapere-Wilczek magnetic monopole for the non-abelian Berry curvature (now allowing for vibrational motion rather than a diatomic of fixed length).   We argue that the resulting approach should be extremely useful for propagating dynamics where angular momentum flows between nuclei and electrons; in particular, if extended to include spin degrees of freedom, the present approach will offer a practical means to study chiral induced spin selectivity through the lens of chiral phonons and coupled nuclear-electronic motion. 
\end{abstract}

\maketitle

\section{Introduction}

Electrons and nuclei routinely exchange and transfer nontrivial amounts of momentum.  For example, over the last two decades, proton coupled electron transfer (PCET\cite{shs:review:pcet}) has become known as a major means to store energy biologically. Given the large mass of a proton, one can expect that a large momentum is exchanged during a PCET process.  More recently, it has been suggested that the chiral induced spin selectivity (CISS) effect reflects angular momentum exchange between electrons and chiral phonons in solids\cite{wu:2020:jpca:spin,fransson:2020prb:vibrational,Zhu2018_ChiralPhonons}, which is an exciting prospect insofar as the nuclei carry a lot of momentum (relative to electrons) and the capacity to unleash that momentum on electrons would be powerful.  Unfortunately, however, our ability to model such momentum transfer remains limited in {\em ab initio} simulations. On the one hand, as far as we are aware, momentum conservation is not enforced for PCET calculations (where the focus is usually energy conservation\cite{shs:pcet:energy_conservation}). On the other hand,  almost all theoretical treatments of  momentum transfer for CISS are typically perturbative in nature\cite{Bloom2024_CISS}. In fact, with the exception of a few calculations on small model systems using exact factorization\cite{Abedi2010_ExactFactor}, we are unaware of a robust, practical means to capture nuclear-electronic momentum transfer non-perturbatively within modern electronic structure approaches.
This momentum transfer problem is also ignored\cite{fatehi:2011:dercouple} by semiclassical surface-hopping methods.\cite{furche:2013:fssh_review}

Fundamentally, the problem is that the approach to molecular quantum problems begins with the Born-Oppenheimer (BO) ansatz, i.e., the ansatz that, to zeroth order, nuclear momentum does not affect the ground states of the electronic wavefunction;  one corrects for the mistakes of BO theory only after generating electronic states, and then usually only perturbatively (e.g., to first order in nuclear velocity).\cite{nafie:1983:jcp:el_momentum,patchkovskii:2012:jcp:electronic_current} Thus, going beyond the BO approximation,\cite{Agostini2022,Takatsuka2015-ky} and making such calculations practical for large molecular systems, contains great value both for chemists and physicists.

In a series of recent works, largely summarized in Ref \cite{Bian2026-cpr}, we have suggested an alternate approach to capture the transfer of  angular and linear momentum between nuclei and electrons, a so-called a semi-classical phase space (PS) electronic structure approach.  Within this ansatz, we are able to include the effects of nuclear motion in the electronic Hamiltonian by approximating   the traditional derivative coupling vector with a one-electron operator $\hbG$ that satisfies the equivalent four symmetries\cite{Bian2026-cpr}:

\begin{figure*}
    \centering
    \includegraphics
    {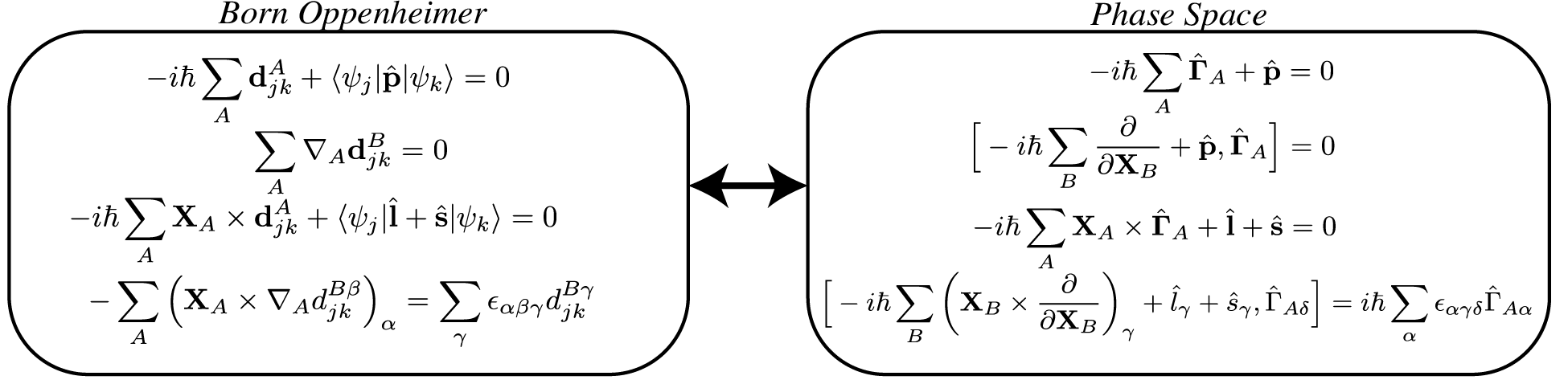}
    \caption{A comparison of the symmetries of the  relevant electron-nuclear couplings within Born- Oppenheimer theory and phase space electronic theory: on the left, we  list the four symmetries of the BO derivative couplings $\bd_{jk}^A$;  on the right, we plot the corresponding symmetries of the corresponding $\hat{\bm \Gamma}$ operator within phase space electronic structure theory. Here, the nuclei are located at positions ${\bm X}_A$, $\hbp$ is the electronic linear momentum, $\hat{\bm l}$  and $\hat{\bm s}$ are the electronic orbital  and spin angular momentum operators, and we index cartesian directions by $\alpha, \beta, \gamma, \ldots$. For this article, we ignore spin entirely (so one can set $\hat{s}$ = 0). } 
 
    \label{fig:dgammasym}
\end{figure*}

These four symmetries\nadine{, as written out for comparison in Figure \ref{fig:dgammasym},} include (i) two phase conventions dictating that an electronic wavefunction should remain unchanged when the same translational or rotational boost is applied to both electrons and nuclei; 
(ii) two statements of invariance in space, dictating that $\hbG$ must not depend on the choice of origin and must transform correctly under a rotation of the overall frame. To satisfy all of these symmetries, we build $\hbG$ as the sum of two sub-components,
$\hbG = \hbG' + \hbG''$,
the first to satisfy translational symmetries and ensure linear momentum conservation; the second to satisfy rotational symmetry and ensure angular momentum conservation.
(Here, we ignore spin and spin-orbit coupling, so there is no need to consider  a third $\hbG'''$ that boosts the spin to the body frame.)

Over the last few years, our ansatz for the best form of such a $\hbG$ operator has undergone a battery of tests: first by comparison to experiment through several different kinds of spectroscopy 
that cannot be explained by BO theory without extensive perturbative treatment\cite{Duston2024_vcd,Tao2026_ROA,Peng2026_lambda,Peng2026_spinrot}
and 
also by comparison to exact quantum mechanical calculations in one-dimension (1D)\cite{Bian2025-vib}.
Regarding the experimental benchmarks, we have now tested PS electronic structure theory as means to recover
vibrational circular dichroism\cite{Duston2024_vcd},
Raman optical activity\cite{Tao2026_ROA},
lambda doubling\cite{Peng2026_lambda}, and 
spin-rotation coupling\cite{Peng2026_spinrot}\nadine{, in both polyatomic and diatomic molecules where appropriate}. 
For all of the spectroscopies above, the major limitation of BO theory is the inability to easily recover electronic angular momentum during the course of nuclear vibrational and/or rotation motion\cite{nafie:1983:jcp:el_momentum,patchkovskii:2012:jcp:electronic_current}. Again, BO theory stipulates that, even when nuclei are moving,  the electronic momentum is always zero--at least for a molecular system with an even number of electrons; the situation is slightly more complex for the odd electron case. (See Refs. \cite{Littlejohn2023,Bian2022_oddBerry,Tao2024_Ehrenfest}for more details and note the on-diagonal derivative coupling is always real in the former case.)  Nevertheless, 
when we go beyond BO theory and implement PS electronic structure theory, we can recover nonzero electronic momentum and therefore all of the signals above can be captured directly. \nadine{Additionally, this picture of non-trivial electronic momentum can also be derived in the solid state, going beyond the traditional conservation of crystal momentum.\cite{Bradbury2026_solids}.}For example, regarding Refs. \cite{Peng2026_lambda} and \cite{Peng2026_spinrot}, we can recover 
the splitting between rotational states from the coupling between nuclear rotation motion and electronic angular momentum (spin and orbital).  Our results exhibit reasonable accuracy for the form of $\hbG$ postulated in Refs. \cite{Tao2025_basis_free,Bradbury2025_SpinCoriolis}.

With regards to the one-dimensional model work in Ref. \cite{Bian2025-vib}, note that for a  small 1D problem, one can easily diagonalize the total Hamiltonian numerically and invent many different, artificial Hamiltonians against which one can benchmark new algorithms and learn quickly (i.e., one can benchmark against many more data points this way than against experiment). As such,
in Ref. \cite{Bian2025-vib}, we 
imagined a system with one heavy particle and one light particle, and we analyzed the overall eigenspectrum and a few electronic observables (especially the electronic momentum) across a range of different potential and heavy/light mass ratios.  Our conclusion was that a PS electronic structure Hamiltonian can strongly outperform BO theory in the limit that the heavy/light mass ratio is not too big. Furthermore,  through comparison against exact calculations, we were also able to more appropriately interpret the dressed operators that arise within PS theory and learn more about the underlying theory of PS electronic structure theory. 

The successes above are very strong endorsements of a PS electronic structure theory approach as a potential replacement for BO theory; at the very least, Refs. \cite{Duston2024_vcd,Tao2026_ROA,Peng2026_lambda,Peng2026_spinrot} strongly indicate that much more research is needed if we seek the best representation of electronic wavefunctions for molecular and material systems.  That being said, one big limitation of the one-dimensional model described above is that the most interesting features of dynamics non-inertial frames cannot be simulated in one dimension. For instance, \nadine{geometric forces, such as the Coriolis and centrifugal forces,} cannot be captured in one-dimension.  More generally, the nature of quantum mechanics is completely different in one dimension than it is in three, where exact calculations must be made in the body frame (rather than lab frame) if one seeks a discrete spectrum; we can expect many technical difficulties to arise because in three dimensions, the angular momentum $\vec{\bm{L}}$ vector operator is not the derivative of any other observable (unlike the linear momentum which satisfies $\vec{\bm p} = m_e \mathrm{d}\vec{\bm x}/\mathrm{d}t$). Nevertheless, the need for three dimensional benchmarks is enormous: without such benchmarks, we are forced to rely on only a handful of experimental comparisons (rather than a plethora of computational benchmarks) to verify our choice of $\hbG.$ 

To that end, the goals of the present article are two-fold. First, motivated by the success of the one-dimensional model in Ref. \cite{Bian2025-vib}, we will establish a general, open-source, efficient GPU framework for diagonalizing a quantum three-body problem with two heavy particles and one light particle occupying two or three dimensions. This code is now publicly available at \cite{shabica_2026_20047795}. Second, using the exact calculation above, we will benchmark the accuracy of a PS electronic structure. We will focus not just on eigenergies but also on angular momentum, analyzing our results for different mass ratios as evaluated on different overall $J$ states.   We focus on different mass ratios because, in so doing, we can lower the electronic gap and study regimes with low-lying electronic states, which are the most interesting  regimes for correlated physics.  Ultimately, we will show that, as in the one-dimensional case, 
a PS semiclassical theory of a minimal molecule not only can recover  excellent vibrational and rotational energy gaps (better than BO theory), but the method can also reproduce all nuclear and electronic observables for the true wavefunction expressed in a body frame, which represents a major success for the method.
While all data presented herein is for three-body systems, we emphasize that phase space electronic structure theory\cite{Tao2025_basis_free} is quite general and can be applied to very large molecular systems.

An outline of this article is as follows. In Sec. \ref{sec:exact:mcm}, we will review the necessary transformations and the resulting equations for diagonalizing the three body problem. We will also describe our implementation. In Sec.\ref{sec:PSeqn}, we will develop the corresponding equations of merit for a PS electronic structure approach. In Sec \ref{sec:BO}, we revisit the Born-Oppenheimer equations for our system. We then compare results for all three theories in Sec. \ref{sec:results}.  In Sec. \ref{sec:MSW}, we show that our PS approach actually generalizes the famous non-abelian curl result found by Moody, Shapere, and Wilczek to apply for a diatomic molecule, \nadine{a third geometric force present only beyond one-dimension}.\cite{Moody1986} In Sec. \ref{sec:conclusion}, we discuss future possibilities for this work, most especially the prospect to include spin in the near future and make contact with experiments involving chiral induced spin separation\cite{naaman:2015:arpc}. 

\begin{figure}
    \centering
    \includegraphics[width=0.65\linewidth]{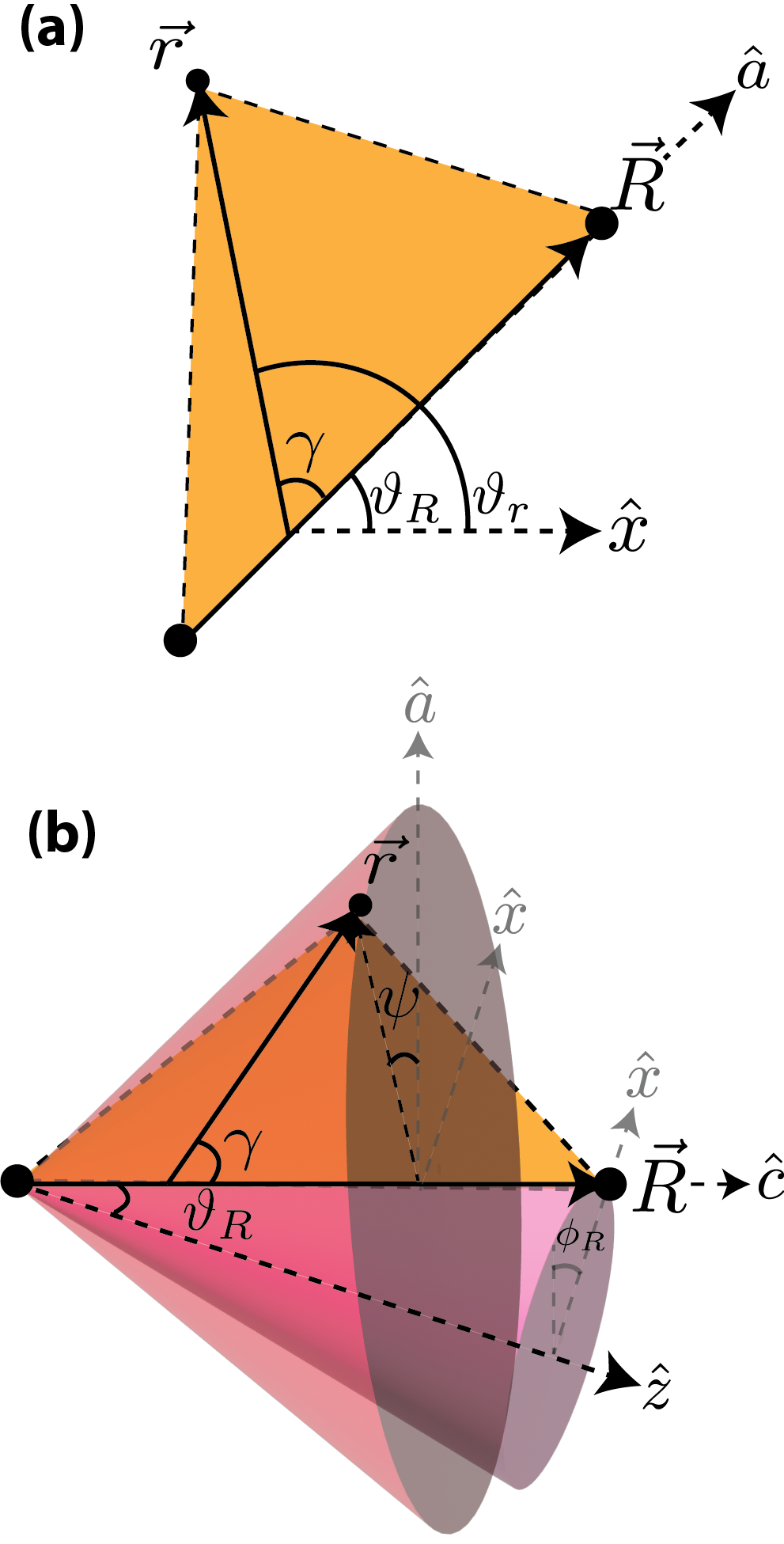}
    \caption{Angles that define the notion of the body-frame in (a) 2-dimensions and (b) 3-dimensions. In (b), we show the cones that define the $\hat{x}\hat{y}-$plane that the molecule can process around in the space-fixed frame according to angle $\phi_R$, and similarly the cone that defines the $\hat{a}\hat{b}$-plane in the body frame where rotation is defined by change in the angle $\psi$.}
    \label{fig:bfc_diagram}
\end{figure}

As a matter of notation, because the focus in this paper is on geometric problems in more than one dimension, below all two and three dimensional vectors and tensors will be written in boldface. Quantum operators are represented with  hats when there is any confusion (within PS and BO theory).
We will not distinguish here between superscripts and subscripts, as they will be used for notational convenience; for example, a vector potential indexed by atoms may be written as $\hbG^I$ or $\hbG_I$  depending on other descriptors, e.g. 
$\hbG'_I$ or $\hbG^I_{jk}$.

\section{The calculation of exact rovibrational energies in 2D and 3D \label{sec:exact:mcm}}
The quantum three-body problem has an extensive history of study---starting with Hylleraas' solution to the $J=0$ problem for the helium atom in 1928-29 \cite{Hylleraas1928,Hylleraas1929} and continuing to the present day.\cite{Breit1930,DattaMajumdar1952,Bhatia1965,Schatz1976_3D,Mukherjee1995,Littlejohn1997,Meremianin2003,Sadhukhan2025-xp} In this article, we  follow most closely the exposition of Ref.\cite{Kuppermann1976_2D,Schatz1976_3D}, which was originally written within the context of atom-diatom scattering, but we shall consider our `light' particle within this context as an electron.

For this section, all observables are quantum operators and we do not include any hats.
We start with a universal Hamiltonian for the 3-body problem in the lab frame, where we imagine having two heavy ('nuclear') particles labeled 1 and 2, and a third light ('electronic') particle.
\begin{equation}
     H = \frac{\uvP_1^2}{2M_1} + \frac{\uvP_2^2}{2M_2} + \frac{\uvp_e^2}{2m_e} + V(\bm{X}_1, \bm{X}_2, \uvr_e) \label{eq:H_lab}
\end{equation}

\subsection{Linear Momentum and Jacobi Coordinates}
As is well known, a transformation to Jacobi coordinates separates the translational degrees of freedom from the translationally invariant (internal) degrees of freedom. Specifically, we make the following change of coordinates: 
\begin{align}
\uvR_{\MCM} &= \frac{1}{M_{\T}}(M_1\uvX_1 + M_2\uvX_2 + m_e\uvr_e) \\
    \uvR &= \uvX_1 - \uvX_2 \\
    \uvr &= \uvr_e - \frac{1}{M_1+M_2}(M_1\uvX_1 + M_2\uvX_2)
\end{align}
Here, $\MCM$ denotes the ``molecular center of mass'', 
\begin{align}
    M_{\T} \equiv M_1+M_2+m_e
\end{align}
is the total mass, and the corresponding Jacobi reduced masses are 
\begin{align}
\mu_R &\equiv M_1M_2/(M_1+M_2) 
\label{eq:muR}
\\
\mu_r &\equiv m_e(M_1+M_2)/M_{\T}   
\label{eq:mur}
\end{align}
A useful equality of the masses is that $M_1 M_2 m_e = M_{\T} \mu_R \mu_r$. The conjugate momenta are given by
\begin{align}
    \uvPC &= \uvP_1+\uvP_2+\uvp_e \\
    \uvPR &= \frac{\mu_R}{M_1} \uvP_1 - \frac{\mu_R}{M_2} \uvP_2 \\
    \uvPr &= \frac{\mu_r}{m_e}\uvp_e - \frac{\mu_r}{M_1+M_2}(\uvP_1 + \uvP_2)
\end{align}
In these new coordinates, the total Hamiltonian is given by
\begin{align}
    H &= \frac{\uvPC^2}{2M_{\T}} + \frac{\uvPR^2}{2\mu_R} + \frac{\uvPr^2}{2\mu_r} + V(\uvR, \uvr) \label{eq:Hmcm}
\end{align}
The Schrödinger equation for the center of mass coordinate can now been separated from the internal Schrödinger equation, making linear momentum conservation obvious. We make a product state wavefunction ansatz.
\begin{equation}
    \Psi_{\TOT}(\uvR,\uvr,\uvR_{\MCM})  = e^{i\bm{K}\cdot \uvR_{\MCM}}\Psi(\uvR,\uvr)
\end{equation}
where {$\hbar\bm{K}$} is the linear momentum of the center of mass. The internal wavefunction $\Psi$ is the corresponding solution to the remaining internal Hamiltonian for $\uvPC = \hbar \bm{K}$. 
Lastly, in accordance  with previous work \cite{Zhang1998-wr}, we will introduce scaled internal vectors (denoted by under-tildes) such that only a single reduced mass appears in the remaining internal Hamiltonian. To that end, we define
\begin{align}
    \mu^2 &= \frac{M_1M_2m_e}{M_{\T}} = \mu_R\mu_r\\
    {\undertilde{\bm{r}}} &= \uvr/a\\
    \undertilde{\bm{R}} &= a \uvR
\end{align}
with $a^2 = \mu_R/\mu = \mu / \mu_r$. 
In these new coordinates, 
 the rescaled and translation reduced Hamiltonian becomes:
\begin{align}\label{eq:Jacobi_coord_nomcm}
    H_{\Int} &= \frac{1}{2\mu}\left(\undertilde{\bm{P}}_R^2 + \undertilde{\bm{P}}_r^2 \right) + V(\undertilde{\bm{R}},\undertilde{ \bm{r}})
\end{align}

Eq. \ref{eq:Jacobi_coord_nomcm} above is valid in any dimension -- one, two or three (hence the bold formatting).

\subsection{Internal Coordinate Transformations and Angular Momentum Conservation in 2D}

The next step is to transform the internal vectors into polar coordinates, such that the total angular momentum can be quantized and  we can eliminate additional degrees of freedom from the internal Hamiltonian. The procedure above is equivalent to going to the body frame of the molecule and is much more straightforward in 2D (relative to 3D) following the methodology given in Ref. \cite{Kuppermann1976_2D}. Dropping the under-tilde notion, we can express
 the internal Hamiltonian in terms of polar coordinates  $R,\, r,\, \vartheta_R,\,\vartheta_r$
 shown in Fig. \ref{fig:bfc_diagram}(a):
\begin{align}\label{eq:lab_frame_int}
    \hat{H}_{\Int}^{\lf} 
    &=\frac{-\hbar^2}{2\mu}\left[ \frac{1}{R}\frac{\partial}{\partial R}R\frac{\partial}{\partial R} + \frac{1}{r}\frac{\partial}{\partial r}r\frac{\partial}{\partial r}  + \frac{1}{R^2}\frac{\partial^2}{\partial \vartheta_R^2} + \frac{1}{r^2}\frac{\partial^2}{\partial \vartheta_r^2}\right] \nonumber \\
      &+ V(R,r,\vartheta_r-\vartheta_R)
\end{align}
The $``\lf"$  superscript in Eq. \ref{eq:lab_frame_int} denotes the fact that the axes are still in the {\em lab frame}.
Let us define the orbital momentum of each Jacobi vector and the total angular momentum of the system as 
\begin{align}
    \hat{l}_r &= \frac{\hbar}{i}\left(\frac{\partial}{\partial \vartheta_r} \right)_{\vartheta_R}\hat{z} \\
    \hat{l}_R &= \frac{\hbar}{i}\left(\frac{\partial}{\partial \vartheta_R} \right)_{\vartheta_r}\hat{z} \\
    \hat{J} &=  \frac{\hbar}{i}\left(\left( \frac{\partial}{\partial \vartheta_r} \right)_{\vartheta_R} + \left(\frac{\partial}{\partial \vartheta_R}\right)_{\vartheta_r}\right)\hat{z}
\end{align}
pointing out of the plane defined by the molecule.
The internal Hamiltonian now becomes: 
\begin{align}
        \hat{H}_{\Int}^{\lf} &= \frac{1}{2\mu}\left[ -\frac{\hbar^2}{R}\frac{\partial}{\partial R}R\frac{\partial}{\partial R} -\frac{\hbar^2}{r}\frac{\partial}{\partial r}r\frac{\partial}{\partial r}  + \frac{\hat{{l}}_R^2}{R^2} + \frac{\hat{{l}}_r^2}{r^2}\right] \nonumber \\
    &+ \hV(R,r,\vartheta_r-\vartheta_R) \label{eq:2d_Hint_lf}
\end{align}

Finally, we replace $\vartheta_r$ by the relative angle $\gamma = \vartheta_r-\vartheta_R$ that leads to a corresponding shift in the definitions of the angular momentum: 
\begin{align}
    \hat{l}_r = \frac{\hbar}{i}\left(\frac{\partial}{\partial \gamma}\right)_{\vartheta_R}\hat{z}\\
    \hat{l}_R = \frac{\hbar}{i}\left( \left(\frac{\partial}{\partial \vartheta_R}\right)_\gamma - \left(\frac{\partial}{\partial \gamma}\right)_{\vartheta_R}\right)\hat{z} \\
    \hat{J} = \frac{\hbar}{i}\left(\frac{\partial}{\partial \vartheta_R}\right)_\gamma \hat{z}.
\end{align}
According to such a change of coordinates, $\vartheta_R$ can  now be interpreted as a rigid rotation of the molecule for a fixed $(R,r,\gamma)$. Again, see Figure \ref{fig:bfc_diagram}(a).
The internal Hamiltonian (now relative to the {\em body frame} [$``\bff"$]) has become:
\begin{align}\label{eq:Hint_unexpanded}
    \hat{H}_{\Int}^{\bff} &= \frac{-\hbar^2}{2\mu}\Bigg[ \frac{1}{R}\frac{\partial}{\partial R}R\frac{\partial}{\partial R} + \frac{1}{r}\frac{\partial}{\partial r}r\frac{\partial}{\partial r}  + \frac{1}{r^2}\frac{\partial^2}{\partial \gamma^2} \nonumber \\
    &+\frac{1}{R^2}\left(\frac{\partial}{\partial \vartheta_R} -\frac{\partial}{\partial \gamma}\right)^2 \Bigg] + \hV(R,r,\gamma)
\end{align}
Lastly, we note that $\vartheta_R$ does not appear in the Hamiltonian; only $\frac{\partial}{\partial \vartheta_R}$ appears.  
Thus, $\hat{J}$ is an operator with a conserved quantum number.  The eigen-functions of $\hat{J}$ are angular functions in the plane of the molecule given by $\phi_J(\vartheta_R) = (2\pi)^{-1/2}e^{iJ\vartheta_R}$ for $J = 0, \pm 1,\pm 2 \cdots$. The decomposition of the total wavefunction as a simultaneous eigenfunction of $\hat{J}$ and $H_{\Int}$ is given by
\begin{equation}
    \Psi_J(R,\vartheta_R, r,\gamma) = \phi_J(\vartheta_R)\zeta_J(R,r,\gamma)
\end{equation}
where $\zeta_J$ is the solution to the reduced body-frame Schrödinger Eq.\ref{eq:2D_exact}.\cite{Kuppermann1976_2D}
\begin{widetext}
\begin{align}
    \hat{H}_J\zeta_J(R,r,\gamma) &= \Bigg[\frac{-\hbar^2}{2\mu}\Bigg(
    \frac{1}{R}\frac{\partial}{\partial R}R \frac{\partial}{\partial R} +
    \frac{1}{r}\frac{\partial}{\partial r}r\frac{\partial}{\partial r}  +
    \left(\frac{1}{r^2} + \frac{1}{R^2} \right)
    \left(\frac{\partial^2}{\partial \gamma^2}\right)- \frac{1}{R^2}\left(J^2 + 2 iJ\frac{\partial}{\partial\gamma}\right)\Bigg)  + \hV(r,R,\gamma)\Bigg]\zeta_J(R,r,\gamma) \label{eq:2D_exact}
\end{align}
\end{widetext}
Note that Eq. \ref{eq:2D_exact} is equivalent to Eq. \ref{eq:Hint_unexpanded} if we replace $-i \hbar \partial/\partial \vartheta_R$ with $\hbar J$ and we now denote $H_{\Int}^{\bff}$ as $H_J$ for a fixed $J$.

\subsection{Internal Coordinate Transformations and Angular Momentum Conservation in 3D \label{exact3D}} 

The transformations above can be extended to three dimensions.  Here, we
closely follow the exposition given in Ref. \cite{Schatz1976_3D}. As in the 2D case, the pair potential $\hV(R,r,\gamma)$ remains a function of the parameters that determine the shape of the molecule. As such, we redefine the $\gamma$ angle as: 
\begin{equation}
    \cos\gamma = \frac{\uvR\cdot\uvr}{|\bm R||\bm r|}
\end{equation} for use in 3D.

At this point, we express the Jacobi vectors in spherical coordinates: $\uvR = (R,\vartheta_R,\phi_R)$ and $\uvr = (r,\vartheta_r,\phi_r)$. 
Here, we use a physicists' convention for spherical coordinates (so that $\vartheta$ is polar and $\phi$ is azimuthal). In these coordinates, the proper form for the Laplacian and internal Hamiltonian $\hat{H}_{\Int}$ is:
\begin{align}
    \hat{H}_{\Int}^{\lf}&=\frac{1}{2\mu}\left[- \frac{\hbar^2}{R^2}\frac{\partial}{\partial R}R^2\frac{\partial}{\partial R} - \frac{\hbar^2}{r^2}\frac{\partial}{\partial r}r^2\frac{\partial}{\partial r}  + \frac{\hat{{\bm l}}_R^2}{R^2} + \frac{\hat{{\bm l}}_r^2}{r^2}\right] \nonumber \\
    &+ \hV(R,r,\gamma)\label{eq:3D_Hint}
\end{align}
Note the similarity between Eq. \ref{eq:2d_Hint_lf} and Eq. \ref{eq:3D_Hint}; the only difference is the Jacobian of the radial coordinates and the definitions of the angular momentum $\hat{\bm{l}}_R$ and $\hat{\bm{l}}_r$ (the latter  angular momentum definitions can be found in just any quantum mechanics textbook\cite{mcquarrie2008quantum,shankar2012principles}).

Next,
we need to define a reference orientation of the molecule, known as the body frame (which we will define by axes $\hat{a} \hat{b} \hat{c}$). We make the definition that $\uvR$ always points in the $\hat{c}$ direction.   
We denote ${Q}$ as the rotational matrix element of $SO(3)$, parametrized in the traditional zyz-convention for the Euler angles, such that ${Q}(\alpha,\beta,\gamma)$ = ${Q}(\phi_R,\vartheta_R,0)$ takes the space-fixed coordinate axes $\hat{x}\hat{y}\hat{z}$ to the axes $\hat{a}\hat{b}\hat{c}$ in the body frame, and such that the rotation aligns the vector $\uvR$ (or equivalently $\hat{c}$) with the $\hat z$-axis.  We further define the so-called `tumbling' angle of the molecule
\begin{equation}
    \tan\psi = \frac{(\bm{r}\times\bm{R})\cdot \hat{a}}{|\bm R||\bm r \cdot\hat{a}|},
\end{equation}
such that a rotation $Q(\phi_R,\vartheta_R,\psi)$ both aligns the $\uvR$-vector with the z-axis and places $\uvr$ in the $\hat{a}\hat{c}$-plane. Variation in $\psi$ therefore corresponds to rotation of the plane of the molecule (or equivalently rotation of the electron) around $\uvR$. These angles and coordinate systems are shown in the diagram in Fig.\ref{fig:bfc_diagram}(b).

For a diatomic molecule---two nuclei and an electron lying in a plane---the shape of the body frame can be parameterized by only two angles; the third angle is extraneous and describes only the motion of the electron. Thus we do not include this third angle when defining the body frame axis, and we work with ${Q}(\phi_R,\vartheta_R,0)$.  With all the foregoing choices,
 the total angular momentum is given by:
\begin{align}
    {\hat{\bm J}} &= {\hat{\bm l}}_R + {\hat{\bm l}}_r \label{eq:3D_Jtot}\\
    \hat{J}_a &= \frac{\hbar}{i}\left( -\csc\vartheta_R \frac{\partial}{\partial \phi_R} + \cot\vartheta_R\frac{\partial}{\partial \psi}\right) \\
    \hat{J}_b &= \frac{\hbar}{i}\frac{\partial}{\partial \vartheta_R} \\
    \hat{J}_c &= \frac{\hbar}{i}\frac{\partial}{\partial\psi},\label{eq:3D_Jz}
\end{align}
as can be found in Ref.~\cite{Schatz1976_3D}.
Note that the total angular momentum around axis $\hat{c}$ (that connects the nuclei) is purely electronic; there is no nuclear component.
The internal Hamiltonian in the body frame is now:
\begin{align}
    \hat{H}_{\Int}^{\bff} 
    &=\frac{1}{2\mu}\Bigg[ -\frac{\hbar^2}{R^2}\frac{\partial}{\partial R}R^2\frac{\partial}{\partial R} - \frac{\hbar^2}{r^2}\frac{\partial}{\partial r}r^2\frac{\partial}{\partial r}  \nonumber \\
    &\hspace{1cm}+ \frac{(\hat{\bm{J}}-\hat{\bm{l}}_r)^2}{R^2} + \frac{{{ \hat{\bm l}}}_r\cdot{\hat{\bm{l}}}_r}{r^2}\Bigg] 
    + \hV(R,r,\gamma)\label{eq:3D_Hint_bf}
\end{align}

Let us denote $\Psi^J_M$ as the total wavefunction (in the lab frame) that diagonalizes $\hat{H}_{\Int}^{\lf}$ and has eigenvalues $\hbar^2J(J+1)$ and $\hbar M$ for operators $J^2$ and $J_z$. 
To convert this wavefunction to the body frame, the relevant coefficients are given by the Wigner-D-matrix($D^J_{M\Omega}$)\cite{Messiah66},
\begin{align}    \Psi^J_M(R,\vartheta_R,&\phi_R,r,\gamma,\psi) \nonumber \\
    &= \sum_\Omega D^{*J}_{M\Omega}(\phi_R,\vartheta_R,0)\Phi^J_\Omega(R,r,\gamma,\psi) \label{eq:ex:bfc_wfc}
\end{align}
Here, $\Phi^J_\Omega(R,r,\gamma,\psi)$ is the body-frame wavefunction, i.e. an eigenfunction of  
 $\hat{H}_{\Int}^{\bff}$,
where  $\hat{J}_c$  (as given in Eq.\ref{eq:3D_Jz}) has eigenvalue 
$\hbar\Omega$ and $\hat{\bm J}^2$ has eigenvalue $\hbar^2 J(J+1)$. Eq. \ref{eq:ex:bfc_wfc} is essentially
Eq. (2.13) of Ref \cite{Schatz1976_3D}, but with a complex conjugation on the D-matrix. The discrepancy arises because those authors used a different definition of the D-matrix than we do; we follow the most common definition nowadays, given, for example,
by Messiah.\cite{Messiah66}

According to Eqs. \ref{eq:3D_Jtot}-\ref{eq:3D_Hint_bf} above, all that remains in order to specify $H_{\Int}^{\bff}$ is an expression for $\hat{l}_r$. As noted above,  since $\uvR$ is aligned with the $\hat{c}$-axis in the body-frame, $\hat{l}_{rc}=\hat{J}_c$.
However, there is still redundancy in the expression for
each channel solution $\Phi^J_\Omega(R,r,\gamma,\psi)$: if we know the eigenvalue $\hbar\Omega$ for $-i \hbar \partial_\psi$, why should we need to further express the wavefunction in terms of $\psi$? Instead,
let us expand the electronic degrees of freedom in terms of a spherical harmonic (the eigenfunction of $\hat{l}^2_r$ and $\hat{l}_{rc})$,

\begin{align} 
\Phi^J_\Omega(R,r,\gamma,\psi)  = \sum_\ell Y_\ell^\Omega(\gamma,\psi)\xi^J_{\ell\Omega}(R,r) \label{eq:3D_bfwfc}
\end{align}

so that
\begin{align} 
\Psi^J_M(R,\vartheta_R,&\phi_R,r,\gamma,\psi) \nonumber \\
    &= \sum_\Omega D^{*J}_{M\Omega}\sum_\ell Y_\ell^\Omega(\gamma,\psi)\xi^J_{\ell\Omega}(R,r)\label{eq:tot_wfc_decomp}
\end{align}
Here,  we can write the other remaining operators of $\hat{\bm{l}}_r$ 
\begin{align}
    \hat{l}_{ra} &= \frac{\hbar}{i}\left(-\cot\gamma\cos\psi \frac{\partial}{\partial\psi} -\sin\psi\frac{\partial}{\partial\gamma} \right) \\
    \hat{l}_{rb} &=\frac{\hbar}{i}\left(-\cot\gamma\sin\psi \frac{\partial}{\partial\psi} +\cos\psi\frac{\partial}{\partial\gamma} \right).
\end{align}
Eq. \ref{eq:tot_wfc_decomp} is akin to the angular/radial separation of the wavefunction done traditionally for the hydrogen atom.
We can identify $\xi^J_{\ell\Omega}(R,r)$ as the angular-coupled radial wavefunction of the molecule, and $D^{*J}_{M\Omega}Y^\Omega_\ell$ is the minimal simultaneous eigenfunction for all angular operators that appear in Eq. \ref{eq:3D_Hint},  The final reduced body-frame Schrodinger equation and its solutions $\xi^J_{\Omega\ell}(R,r)$ can
be found by inserting $\Phi^J_\Omega$ from Eq. \ref{eq:3D_bfwfc} into Eq. \ref{eq:3D_Hint_bf}, multiplying by $\left(Y_{\ell'}^{\Omega'}\right)^*$ and integrating over angles $\gamma$ and $\psi$.
The solution is \cite{Schatz1976_3D}:
\begin{widetext}
\begin{align}
    H_J \xi^J_{\ell\Omega}(R,r) &= t^{J\ell}_{\Omega\Omega}\xi^J_{\ell\Omega}(R,r) +  t^{J\ell}_{\Omega \Omega+1}\xi^J_{\ell\Omega+1}(R,r) + t^{J\ell}_{\Omega\Omega-1}\xi^J_{\ell\Omega-1}(R,r) + \sum_{\ell'} \langle \ell'\Omega |\hV(r,R,\gamma)|\ell\Omega\rangle \xi^J_{\ell'\Omega}(R,r), \label{eq:3D:Hxi}\\
     t^{Jj}_{\Omega\Omega} &= \frac{\hbar^2}{2\mu}\left( -\frac{1}{R^2}\frac{\partial}{\partial R} R^2\frac{\partial}{\partial R} -\frac{1}{r^2}\frac{\partial}{\partial r} r^2\frac{\partial}{\partial r} + \frac{\ell(\ell+1)}{r^2} + \frac{1}{R^2}\left(J(J+1) -2\Omega^2 + \ell(\ell+1)\right) \right),\label{eq:3D:Hxi_diag}\\ 
     t^{Jj}_{\Omega \Omega\pm 1} &= -\frac{\hbar^2}{2\mu R^2}\sqrt{J(J+1) - \Omega(\Omega\pm1)}\sqrt{\ell(\ell+1)-\Omega(\Omega\pm1)} \label{eq:3D:Hxi_offdiag}, \hspace{2cm} \ell \geq |\Omega| \hspace{0.2cm}\text{and}\hspace{0.2cm} J \geq |\Omega|
\end{align}
\end{widetext}
The potential in the spherical harmonic basis can be written with the trivial integral over $\psi$ computed as,\footnote{$Y_\ell^m(\gamma,\psi) = \frac{1}{\sqrt{2\pi}}e^{im\psi}\mathcal{P}^m_\ell(\gamma)$, with $\mathcal{P}^m_\ell$ is the normalized associated Legendre polynomial, $\mathcal{P}^m_\ell(\gamma) = f_m \sqrt{\frac{2\ell+1(\ell-|m|)!}{2(\ell+|m|)!}}P^{|m|}_\ell(\cos\gamma)$, and $f_m$ is the Condon-Shortley phase convention $f_m = 1,\, m \leq0$ and $f_m = (-1)^m, \, m>0$}
\begin{align}
    \bra{\ell'\Omega}&\hV(R,r,\gamma)\ket{\ell\Omega} \nonumber \\
    &= \int_0^{\pi} d\gamma \sin(\gamma) \mathcal{P}_{\ell'}^{\Omega}(\gamma) \mathcal{P}_\ell^{\Omega}(\gamma) \hV(r,R,\gamma) \label{eq:Vsph}
\end{align}
for $\ell,\ell' > |\Omega|$ and where $\mathcal{P}^\Omega_\ell(\gamma)$ is the normalized associated Legendre polynomial of $\gamma$ with traditional Condon-Shortley normalization.

The solutions to the body-frame Hamiltonian given in \ref{eq:2D_exact} and \ref{eq:3D:Hxi} are found iteratively using Davidson's algorithm\cite{Davidson1975}. More details of the numerical solutions are given in Appendix \ref{appendix:numerics}.

\section{The 3-body problem where we separate light and heavy particles }\label{sec:PSeqn}

\subsection{The Born-Oppenheimer or Born-Huang Framework}
Before addressing the three-body problem within phase space electronic structure theory, we begin with the Born-Oppenheimer perspective.
 In this framework, we begin by partitioning the total Hamiltonian given in Eq. \ref{eq:H_lab} into electronic and nuclear terms, and denote $\hat U_{\el}$ as the unitary that diagonalizes $\hat{H}_{\el}$,  
\begin{align}
     \hat{H} &= \frac{\hat{\uvP}_1^2}{2M_1} + \frac{\hat{\uvP}_2^2}{2M_2} + \hat{H}_{\el} \label{eq:PSHfull}\\
    \hat{H}_{\el} &= \frac{\hat{\uvp}_e^2}{2m_e} + \hat{V}(\hat{\uvX}_1, \hat{\uvX}_2, \hat{\uvr}_e) \label{eq:H_el}\\
    \left(\hat{H}_{\ad}\right)_{jk} & = 
    (\hat{U}_{\el}^\dagger \hat{H} \hat{U}_{\el})_{jk} \\
    &= \sum_{A=1,2} \sum_l \frac{1}{2M_A}\left(\hat{\uvP}_A\delta_{jl} -i\hbar \bm{d}^A_{jl}\right)\nonumber \\
    &\qquad \cdot\left(\hat{\uvP}_A\delta_{lk} -i\hbar \bm{d}^A_{lk}\right) + E^{\el}_j \delta_{jk} \label{eq:H_adiabatic_full}
\end{align}
Here, we have denoted the adiabatic electronic eigenstates by $j,k,l$ so that $\hat{H}^{\el} \ket{\psi_j} = E_j^{\el} \ket{\psi_j}$ and Eq. \ref{eq:H_adiabatic_full} is a matrix of nuclear operators (in the basis of adiabatic electronic eigenvectors).
The non-adiabatic couplings $\bm{d}^A_{lk}$ in Eq. \ref{eq:H_adiabatic_full}  are  given by $\bm{d}^A_{lk} = \left< \psi_l \middle| \hat{\nabla}_{\uvX_A} \psi_k\right>$. 

In Eq. \ref{eq:PSHfull}, one finds a sum over two atoms and one electron. In other words, we have not eliminated the overall center of mass motion. 
Now whereas in the exact solution,  Eq. \ref{eq:Hmcm} above, one separates out the total (nuclear plus electronic center of mass), within BO theory, it is natural to separate out the {\em nuclear} center of mass and change to nuclear Jacobi coordinates:

\begin{align}
    \uvR_{\NCM} &= \frac{M_1\uvX_1 + M_2\uvX_2}{M_1+M_2}\label{eq:R_NCM} \\
    \uvR &= \uvX_1 - \uvX_2 \label{eq:R12}\\
    \uvP_{\NCM} &= \uvP_1 +\uvP_2 \label{eq:PNCM}\\
    \uvPR &= \frac{\mu_R}{M_1}\uvP_1 -\frac{\mu_R}{M_2}\uvP_2 \label{eq:PR}
\end{align}
Here, $\NCM$ denotes the  ``nuclear center of mass''. In these new coordinates, the final BO Hamiltonian becomes
\begin{align}
    \left(\hat{H}_{\ad}\right)_{jk} 
    &= \sum_l\Bigg[\frac{(\hat{\uvP}_{\NCM}\delta_{jl}- {\bm{p}}_{e,jl})(\hat{\uvP}_{\NCM}\delta_{lk}-{\bm{p}}_{e,lk})}{2(M_1+M_2)} \nonumber \\
    &+\frac{(\hat{\uvP}_R\delta_{jl}-i\hbar {\bd}^R_{jl})(\hat{\uvP}_R\delta_{lk}-i\hbar {\bd}^R_{lk})}{2\mu_R}\Bigg]+ E^{\el}_j \delta_{jk} \label{eq:H_adiabatic}
\end{align}

In Eq. \ref{eq:H_adiabatic} above, the derivative coupling for the new $\bR$ coordinate is: $\bm d^R_{jl} = \left< \psi_j \middle| \hat{\nabla}_{\uvR} \psi_l\right> = \frac{\mu_R}{M_1}\bm d^1_{jl} - \frac{\mu_R}{M_2}\bm d^2_{jl}$. For the $\bR_{\NCM}$ coordinate, we find the appearance of ${\bm p}^e_{jl} = \bra{\psi_j}\hat{\uvp}_e\ket{\psi_l}$, which arises naturally through the translational invariance of $\bm d^A_{jl}$; in other words, the equality  $\sum_A -i \hbar \bm d^A_{jl} + \hat{\bm p}^e_{jl} = 0$ \cite{Littlejohn2023, yanzewu:2024:jcp:pssh_conserve, Littlejohn2024} leads to the fact that
\begin{align}
    i\hbar\bd^{\NCM}_{jl} =  {\hat{\bm p}}^e_{jl}.
    \label{eq:dNCM:pe}
\end{align}
The term proportional to $\hat{\uvp}_e\!^2/2(M_1+M_2)$ in Eq. \ref{eq:H_adiabatic} is known as the mass-polarization term\cite{jensen}.

Note that thus far, everything in this section has been quantum mechanically exact---Eq.~\ref{eq:H_adiabatic} still contains nuclear operators. 
One means to take the classical limit of a  quantum operator is to apply a Wigner  transformation, 
\begin{eqnarray}
    O_W(\bm{P},\bm{X}) &=& \int d\uvX'e^{-\frac{i}{\hbar}\bm{P}\cdot \uvX'} \bra{\uvX + \frac{\uvX'}{2} }\hat{O} \ket{ \uvX - \frac{\uvX'}{2}} \nonumber\\\label{eq:wigner}
\end{eqnarray}
whose inverse is a Weyl transformation:
\begin{equation}
\bra{\uvX}\hat{O}\ket{\uvX'} = \int \frac{d\bvP}{(2\pi\hbar)^d} e^{\frac{i}{\hbar}\bvP\cdot(\uvX-\uvX')} O_W\left(\frac{\uvX +\uvX' }{2},\bvP\right)\label{eq:Weyl_gen},
\end{equation}
where $d$ denotes the number of spatial dimensions.
Taking the Wigner transform of the nuclear operators for each $jk$  matrix element in Eq. \ref{eq:H_adiabatic_full} yields:
\begin{eqnarray}
    H^{\PS}_{jk}(\bm{X},\bm{P})
    &=& \sum_{A=1,2} \sum_l \frac{1}{2M_A}\left({\uvP}_A\delta_{jl} -i\hbar \bm{d}^A_{jl}\right)\nonumber \\
    & & \qquad \cdot\left({\uvP}_A\delta_{lk} -i\hbar \bm{d}^A_{lk}\right) + E^{\el}_j \delta_{jk} \label{eq:H_adiabatic_full2}
\end{eqnarray} 

We can then construct an electronic operator\footnote{Note that Eq. \ref{eq:H_adiabatic_full3} is not the Wigner transform of the total nuclear-electronic Hamiltonian. As noted in Ref. \cite{izmaylov:2014:qcle_berry}, first converting to the adiabatic representation and then taking the Wigner transform is not equivalent to  first  taking the Wigner transform and then converting to the adiabatic representation.} suited to this basis:
\begin{widetext}
    \begin{eqnarray}
    \hat{H}^{\PS}(\bm{X},\bm{P}) = \sum_{jk}
    \ket{j}
    \left(
    \sum_{A=1,2} \sum_l \frac{1}{2M_A}
     \left({\uvP}_A\delta_{jl} -i\hbar \bm{d}^A_{jl}\right) \cdot\left({\uvP}_A\delta_{lk} -i\hbar \bm{d}^A_{lk}\right) + H^{\el}_{jk} \right) \bra{k}\label{eq:H_adiabatic_full3}
\end{eqnarray} 
\end{widetext}

Note that we have removed the hat from the nuclear momentum ($\hbP$) operator in Eqs. \ref{eq:H_adiabatic_full2} and \ref{eq:H_adiabatic_full3} -- after a Wigner transform to phase space, $\bP$ is a symbol not an operator. Eq. \ref{eq:H_adiabatic_full3} is Shenvi's phase space electronic Hamiltonian\cite{Shenvi2009-jcp}, which has  recently been studied by Polkovnikov and co-workers.\cite{polkovnikov:2026:pnas}

Working with the Wigner transformation usually makes it {\em harder} to find the exact solution for a quantum mechanical problem, but the transformation does make it easier to find new approximate solutions, as we now describe.

\subsection{A Phase Space Electronic Hamiltonian}

\subsubsection{The Basic Framework}
Taking inspiration from Shenvi's superadiabatic phase space surface hopping formalism\cite{Shenvi2009-jcp}, 
the essence of a phase space electronic Hamiltonian is to work with an electronic operator like Eq. \ref{eq:H_adiabatic_full3} above (with classical nuclear degrees of freedom) rather than with the standard Born-Oppenheimer electronic Hamiltonian $\hat{H}_{\el}$.

 That being said, one does not want to work directly with the true nonadiabatic coupling ${\bm d}_{jk}^A$ for several reasons. First, the nonadiabatic coupling diverges near state crossings, which render the resulting potential energy surfaces unphysical and unstable. Indeed, Izmaylov {\em et al}\cite{izmaylov:2014:qcle_berry} have warned against using Eq. \ref{eq:H_adiabatic_full2} as a starting point for quantum-classical Liouville equation (QCLE)\cite{kapral:1999:jcp} calculations.  Second, the nonadiabatic couplings are not uniquely defined for Hamiltonia exhibiting degeneracy, and thus the resulting framework would not be well defined for problems with spin (an especially interesting area\cite{wu:2022:pssh}).
Third, calculating derivative couplings and then differentiating them for dynamics would be prohibitively expensive.
To that end, within modern phase space electronic structure theory, 
we approximate matrix elements of the non-adiabatic coupling vector ${\bm d}_{jk}^A$ by the matrix elements of  a simple one-electron operator $\hat{\bm \Gamma}_A$ (with matrix elements ${\hbG}_{jk}^A$ in the adiabatic basis);  the one-electron operator $\hbG_A$ is constructed from symmetry considerations (see Fig. \ref{fig:dgammasym})
 and can be expressed as a function of $\hbr$ and $\hbp$ (see Eqs. \ref{eq:etf_define}  and \ref{eq:erf_define} below). Mathematically, we posit an electronic Hamiltonian of the form (compare against Eq. \ref{eq:H_adiabatic_full3}): 
\begin{eqnarray}
    \hat{H}_{\PS}(\bm{X},\bP) = \sum_{A=1,2} \frac{(\bP_A - i\hbar\hbG_A({\bm X}))^2}{2M_A}
    + \hat{H}_{\el}({\bm X})
    \label{eq:gamma_general}
\end{eqnarray}
Within QCLE calculations, a similar preconditioning was suggested previously by Wu {\em et al.}\cite{yanze:2023:jcp:pqcle}

Eq. \ref{eq:gamma_general} is non-standard in chemical and condensed matter physics, and one might worry about the choice of $\hbG$  and whether or not one can use the  electronic  basis of $\hat{H}_{\PS}$ as a starting point for a rigorous calculation (as is possible for BO theory). To that end, Ref. \cite{Wu2025} indeed demonstrates how to rigorously use this PS reference framework to diagonalize the total nuclear-electronic Hamliltonian and extract exact eigenstates using Littlejohn-Flynn theory\cite{littlejohn:flynn:1991:pra:coriolis}. Thus, in principle, one can build a PS electronic structure Hamiltonian from just about any choice of $\hbG$; that being said, however, convergence to the exact solution is of course faster for an optimized $\hbG$ operator that captures the correct physics and the correct symmetries (a fact which has motivated us to continuously search for improved $\hbG$ operators).

Now, importantly, coordinate transformations can be performed for phase space electronic Hamiltonians just as for the BO electronic Hamiltonians.  To that end, we  simply insist that the $\hbG$ operators transform as momenta. In other words, mirroring the transformation  of $\bP$ in Eqs. \ref{eq:PNCM}-\ref{eq:PR}, we define:

\begin{eqnarray}
\label{eq:gamma_NCM}
\hat{\bm \Gamma}_{\NCM} &=& 
    \hat{\bm \Gamma}_{1} + 
    \hat{\bm \Gamma}_{2}\\
    \hat{\bm \Gamma}_R &=& \frac{\mu_R}{M_1}\hat{\bm \Gamma}_1 - \frac{\mu_R}{M_2}\hat{\bm \Gamma}_2
    \label{eq:gamma_rel}
\end{eqnarray}
In these new coordinates, Eq. \ref{eq:gamma_general} above becomes:
\begin{eqnarray}
    \hat{H}_{\PS} =  \frac{(\bP_{\NCM} - i\hbar\hbG_{\NCM})^2}{2(M_1 + M_2)} + 
    \frac{(\bP_R - i\hbar \hbG_R)^2}{2\mu_R}
    + \hat{H}_{\el}(\bR)
    \nonumber\\
    \label{eq:HPS_rel_coord}
\end{eqnarray}

As a sidenote, in all that follows, we will 
set the electronic origin to be $\bm{R}_{\NCM}$, such that for the electronic operators in $\hat{H}_{\el}(\bR)$, we replace $\hat{\uvr}_e \to \hat{\uvr}_e -\bm{R}_{\NCM}$.  This procedure is not entirely equivalent to the more standard approach by which  the center of mass is removed (as in Sec. \ref{sec:exact:mcm} above), but the overall effect is the same if we requantize and then diagonalize the final Hamiltonian.\cite{footnote_ncm_mcm} (The same choice for  electronic origin will also be made for the BO calculations in Sec. \ref{sec:BO}.)

\subsubsection{Electron translation and rotation factors}\label{sec:etferf}

In practice, for a diatomic molecule without spin, following Appendix   \ref{appendix:gamma''} (and Ref.  \cite{Tao2025_basis_free}), the form of $\hat{\bm \Gamma}_1 = \hat{\bm \Gamma}_1' + \hat{\bm \Gamma}_1''$ for atom $1$ is broken into two terms. The first term $\hat{\bm \Gamma}'_1$ allows electronic linear momentum to ``talk'' to nuclear linear momentum; the second term $\hbG''_1$  allows the electronic angular momentum to ``talk'' to the nuclear angular momentum. Each of these functions depends on a global space partitioning function,
\begin{align}
    \hat{\Theta}_1(\hat{\uvr};\uvX_1,\uvX_2) = \frac{e^{-|\hat{\uvr}-\uvX_1|^2/\sigma^2}}{e^{-|\hat{\uvr}-\uvX_1|^2/\sigma^2} +e^{-|\hat{\uvr}-\uvX_2|^2/\sigma^2}} \label{eq:Theta}
\end{align}
and analogously for $\hat{\Theta}_2$ such that $\hat{\Theta}_1 + \hat{\Theta}_2 = 1$ for all $\hat{\uvr}$. The broadening parameter $\sigma = 1 $Bohr is chosen to be appropriate for the length scale of the typical bond.

We then define 
\begin{align}
    \hat{\bm \Gamma}'_1 = \frac{-i}{2\hbar}\left\{\hat{\Theta}_1, \hat{\uvp}_e\right\}\label{eq:etf_define}
\end{align}
and analogously for $\hbG'_2$.
Here $\left\{\hat{A},\hat{B}\right\}=\hat{A}\hat{B}+\hat{B}\hat{A}$ is the anticommutator.

Simplifying Appendix \ref{appendix:gamma''} for the diatomic case, the  $\hbG''$ term is  defined via 
\begin{align}
    \hat{\bm \Gamma}''_1 &= -M_1\left(\uvX_1 \times ({I}^{-1}\hat{\bm J}_1) +\uvX_1\times({I}^{-1}\hat{\bm J_2})\right) 
    \label{eq:erf_define}
    \\
    \hat{\bm J}_1 &= \frac{-i}{2\hbar}(\hat{\uvr}-{\uvX}_1)\times (\hat{\Theta}_1\hat{\uvp}_e + \hat{\uvp}_e\hat{\Theta}_1)\label{eq:erf_J1_define} \\
    \hat{\bm J}_2 &= \frac{-i}{2\hbar}(\hat{\uvr}-{\uvX}_2)\times (\hat{\Theta}_2\hat{\uvp}_e + \hat{\uvp}_e\hat{\Theta}_2) 
\label{eq:erf_J2_define} 
\end{align}
and analogously for $\hbG''_2$.
Here,  ${I}^{-1}$ is the inverse moment of inertia tensor, adapted to handle the degeneracy of the $\bvR$ axis.

The definitions in Eqs. \ref{eq:etf_define} - \ref{eq:erf_define} automatically satisfy the constraints established in Fig. \ref{fig:dgammasym}.  Moreover, 
\begin{eqnarray}
    \hbG_{\NCM}= \hbp_e/(i \hbar)
\end{eqnarray}
so that Eq. \ref{eq:HPS_rel_coord} 
automatically recovers the mass-polarization term mentioned above (below Eq. \ref{eq:dNCM:pe}).

\subsubsection{The Second Order Term}\label{sec:gammasq}

One of the most interesting facets of the PS electronic Hamiltonian in Eq. \ref{eq:gamma_general} is the second order term, $-\hbar^2\sum_A \hbG_A \cdot \hbG_A/(2M_A)$. In general,  this term is smaller than the $\bP \cdot \hbG$ term and thus has often been excluded from past phase space calculations\cite{Bian2025-vib,Duston2024_vcd,Tao2026_ROA,Peng2026_lambda,Peng2026_spinrot}. That being said, we will show below that these terms can be crucial when the mass difference from nucleus to electron is made artificially small. Moreover, while the present manuscript focuses on a diatomic system with only one electron, one of the outcomes of the analysis above is the realization that the $-\hbar^2\sum_A \hbG_A \cdot \hbG_A/(2M_A)$ term is a two-electron term that causes electron correlation.  After all, the mass-polarization term is a two-electron term, and this term is captured exactly by the $\hbG_{\NCM} \cdot \hbG_{\NCM}/(2(M_1+M_2))$ term; it stands to reason that the other second order terms should also be two-electron terms. Thus, there was clearly an error in Ref. \cite{Tao2025_basis_free,Bian2026-cpr} where we hypothesized that $\sum_A \hbG_A \cdot \hbG_A/(2M_A)$ might be a one-electron operator; for maximal accuracy, future phase space work will need to treat both the Coulomb operator and the second order $-\hbar^2\sum_A \hbG_A \cdot \hbG_A/(2M_A)$ term as two-electron operators.

\subsubsection{Interpretation of the nuclei and choice of body frame \label{sec:PS_Ptheta}}

Within the exact calculations from Sec. \ref{sec:exact:mcm}, one will find rovibrational energies that depend on the quantum number $J$. 
Thus, in order to compare exact results against  a phase space electronic structure calculation (or a BO calculation), one requires a conserved nuclear observable within phase space electronic structure theory that maps to  $J$ semiclassically. 
Now, according to a thorough examination of BO theory, one learns that the nuclear momentum $\bP_{NCM}$ in Eq. \ref{eq:H_adiabatic} in fact represents the total internal linear momentum\cite{Littlejohn2024}, and the nuclear angular momentum is representative of the total angular momentum.\cite{Littlejohn2023} Indeed, this realization is at the heart of phase space electronic structure theory.

In two dimensions, the mapping is clear: if we identify $P_\vartheta$ as the component of the momentum tangent to the inter-nuclear axis, then $P_\vartheta/\hbar$ is the classical analog of the total angular momentum quantum number $J$. This mapping is especially easy to see with the molecule oriented such that $\vartheta=0$ when the Cartesian $\bvP$ can be expressed $\bvP = (P_R,P_\vartheta/R)$.
Thus, for the two-dimensional final hamiltonians and results in Appendix \ref{appendix:2D_data}, when we compare against exact results for angular momentum $J$, we set $\bP_\vartheta/\hbar = J$ for the BO and/or phase space electronic structure results.
While any phase space (or Born Oppenheimer) electronic structure calculation will usually be defined with the molecule in some reference orientation, all of the resulting energies are independent of this choice of reference orientation.

For 3D simulations,  the analogy between the angular part of $\bvP$ and the corresponding quantum number $J$ is less obvious, but it will be helpful to express the coordinates of $\bvR$ and $\bvP_R$ in spherical coordinates but along cartesian axes.
In terms of the coordinates $(R,\mathbf{\vartheta}_R,\mathbf{\phi}_R)$ and $(P_R, P_\vartheta, P_\phi)$, 
we can express $\bvR$ and $\bvP$ in a cartesian frame as:
\begin{widetext}
\begin{align}
    \bR &=  \begin{pmatrix}
        R_x \\ R_y \\ R_z 
    \end{pmatrix} = R \begin{pmatrix}
        \sin\vartheta_R\cos\phi_R \\
        \sin\vartheta_R\sin\phi_R \\
        \cos\vartheta_R
    \end{pmatrix}\\
    \bvP_{R} &= \begin{pmatrix}
        P_x \\ P_y \\ P_z
    \end{pmatrix}=\begin{pmatrix}  P_R\sin\mathbf{\vartheta}_R\cos\mathbf{\phi}_R+ \frac{P_\vartheta}{R}\cos\mathbf{\vartheta}_R\cos\mathbf{\phi}_R - \frac{P_\phi}{R\sin\mathbf{\theta}_R}\sin\mathbf{\phi}_R \\
     P_R\sin\vartheta_R\sin\phi_R  + \frac{P_\vartheta}{R}\cos\vartheta_R\sin\phi_R+\frac{P_\phi}{R\sin\vartheta_R}\cos\phi_R \\
     P_R\cos\vartheta_R-\frac{P_\vartheta}{R}\sin\vartheta_R 
     \end{pmatrix} \label{eq:pnew}
\end{align}
\end{widetext}
As is well known for the two-body problem, the nuclear angular momentum is then  represented by \cite{Bunker2006}
\begin{align}
    {\hat{\bm l}}_R^2 &= P_\vartheta^2 + \frac{P_\phi^2}{\sin^2\mathbf{\vartheta}_R} \label{eq:PS_Ptheta_J} 
\end{align}
where $\hat{R}$ now indicates the direction of $\bR.$

Now, it would seem natural  to force the eigenvalues of     ${\hat{\bm l}}_R^2$    to be equal to the total angular momentum eigenvalue $\hbar^2 J(J+1)$
\begin{align}
    \hbar^2 J(J+1) &\approx P_\vartheta^2 + \frac{P_\phi^2}{\sin^2\mathbf{\vartheta}_R}
\end{align}
That being said, this rule of thumb (with only one equation) is not enough to specify both phase space variables, $P_\vartheta$ and $P_{\phi}$. 
Of course, one cannot define these momenta without also addressing their conjugate angles $\vartheta_R$ and $\phi_R$, which define the body frame, which we address in Sec. \ref{sec:PSbodyframe} below.

\subsubsection{The Form of the Phase Space Electronic Structure Hamiltonian\label{sec:psdefinefull}}

Using Eq. \ref{eq:pnew} above, 
the Hamiltonian takes the form:

\begin{widetext}
\begin{eqnarray}
\hH_{\PS}(P_R,P_\vartheta,P_\phi,R,\vartheta_R,\phi_R)
&=&\frac {\left(\bvP_{R}- i\hbar \hat {\bm \Gamma}_R\right)^2} {2\mu_{R}}
+ \hat H_{\el}(\hat x,\hat y,\hat z)
\\&=& \frac{1}{2\mu_{R}}\left(P_R^2+\frac{P_\vartheta^2}{R^2}+ \frac{P_\phi^2}{R^2\sin^2\vartheta}\right)\nonumber\\&&-\frac{i\hbar}{\mu_{R}}\hat{\Gamma}_x \cdot \left(P_R\sin\mathbf{\vartheta}_R\cos\mathbf{\phi}_R+ \frac{P_\vartheta}{R}\cos\mathbf{\vartheta}_R\cos\mathbf{\phi}_R - \frac{P_\phi}{R\sin\mathbf{\theta}_R}\sin\mathbf{\phi}_R\right)\nonumber\\&&-\frac{i\hbar}{\mu_{R}}\hat{\Gamma}_y \cdot \left(P_R\sin\vartheta_R\sin\phi_R  + \frac{P_\vartheta}{R}\cos\vartheta_R\sin\phi_R+\frac{P_\phi}{R\sin\vartheta_R}\cos\phi_R\right)\nonumber\\&&-\frac{i\hbar}{\mu_{R}}\hat{\Gamma}_z \cdot \left(P_R\cos\vartheta_R-\frac{P_\vartheta}{R}\sin\vartheta_R \right)\nonumber\\
&&-\frac{\hbar^2}{2\mu_{R}}\left(\hat{\Gamma}_x^2 + \hat{\Gamma}_y^2 + \hat{\Gamma}_z^2\right) + \hat H_{\el}(\hat x,\hat y,\hat z)
\end{eqnarray}
\end{widetext}
Here, 
$\hbG_x$ is the $\hat{x}$ component of $\hbG_R$;
here and below,
we will  drop the $R$-subscript on $\hbG$.
The explicit form of the $\hbG$ operators will be defined below in Eqs. \ref{eq:gx_derive}-\ref{eq:gz_derive}. Lastly, we note the $\hbG_{\NCM} \cdot \hbG_{\NCM}/(2(M_1+M_2)) = \hbp^2/(2(M_1+M_2)$ term has been absorbed into the electronic Hamiltonian by using an kinetic energy with the correct reduced mass $\hat{\uvp}_e^2/2\mu_r$ as also appears in the exact calculation.

\subsubsection{The Final Equations of Motion: Choice of the Body Frame in 3D}\label{sec:PSbodyframe}

The choice of $\vartheta_R$ and $\phi_R$ for a PS electronic structure calculation is analogous to specifying a body frame in an exact calculation.
For 3D, choosing 
$\vartheta_R$ and $\phi_R$
is more involved than in 2D.
The PS equations simplify  most when $\bvR$ is aligned with the $x$-axis, namely when $\vartheta_R = \frac{\pi}{2}$ and $\phi_R = 0$. 
Under this choice of frame,
Eq. \ref{eq:PS_Ptheta_J} is non-singular and
$\bvP_R = (P_R, P_\phi/R, -P_\vartheta/R)$ in a cartesian  $xyz$ frame of axes. In this orientation, $P_\phi$ corresponds to a nuclear rotation around the $\hat{x}$-axis  (the degenerate axis of the nuclei with $C_\infty$ symmetry in a chemist's mind), an operation which has corresponding eigenvalue $\Omega$ in the exact calculation. $P_\vartheta$ corresponds to a rotation around a fixed $\hat{y}$-axis. 
$P_\phi$ can be set to 0 without loss of generality, so that:
\begin{eqnarray}
    P_\phi &=& 0 \\
    P_\vartheta &=& \hbar \sqrt{J(J+1)}
\end{eqnarray}

In this frame, the moment of inertia matrix given in  coordinates $xyz$ is
\begin{align}
    I &= \mu_R R^2 \begin{pmatrix}
        0 & 0 & 0 \\
        0 & 1 & 0 \\
        0 & 0 & 1 \\
    \end{pmatrix}\label{eq:mom_inertia}
\end{align}
The inverse is then given by the inverting only the non-singular part of $I$.
\begin{align}
    I^{-1} &= \frac{1}{\mu_R R^2} \begin{pmatrix}
        0 & 0 & 0 \\
        0 & 1 & 0 \\
        0 & 0 & 1 \\
    \end{pmatrix}\label{eq:Iinv}
\end{align}
In this frame, $\hH_{\PS}$ simplifies to
\begin{widetext}
    \begin{eqnarray}
        \hH_{\PS}(P_R,P_\vartheta,P_\phi,R,\frac{\pi}{2},0)  &=& \frac{1}{2\mu_{R}}\left(P_R^2+\left(\frac{P_\phi}{R}\right)^2 + \left(\frac{P_\vartheta}{R}\right)^2\right) - \frac{i\hbar}{\mu_{R}}\Big(P_R \hat \Gamma_x + \frac{P_\phi}{R}\hat \Gamma_y -\frac{P_\vartheta}{R}  \hat \Gamma_z\Big)\nonumber\\
    &-&\frac{\hbar^2}{2\mu_{R}}\left(\hat{\Gamma}_x^2 + \hat{\Gamma}_y^2 + \hat{\Gamma}_z^2\right) + \hat H_{\el}(\hat x,\hat y,\hat z) \label{eq:HPS_final}
    \end{eqnarray}
or even more explicitly for a calculation with $J$ angular momentum (under the choice $P_\phi=0$): 
    \begin{eqnarray}
        \hH_{\PS}^J(P_R,R)  &\equiv& \frac{1}{2\mu_{R}}\Big(P_R^2 + \frac{\hbar^2 J(J+1)}{R^2}\Big) - \frac{i\hbar}{\mu_{R}}\Big(P_R \hat \Gamma_x  -\frac{\hbar \sqrt{J(J+1)}}{R}  \hat \Gamma_z\Big)\nonumber\\
    &-&\frac{\hbar^2}{2\mu_{R}}\left(\hat{\Gamma}_x^2 + \hat{\Gamma}_y^2 + \hat{\Gamma}_z^2\right) + \hat H_{\el}(\hat x,\hat y,\hat z) 
    \label{eq:compare:bunker}
    \end{eqnarray}
Above, following Sec. \ref{sec:psdefinefull}, $(\hG_x,\hG_y,\hG_z)$ are defined as:
\begin{eqnarray}
\label{eq:gx_derive}
    \hat{\Gamma}_x &=&  \hat{\Gamma}'_x + \hat{\Gamma}''_x
    = \frac{1}{2i\hbar}
    \Big\{\hat{\Theta}_R,\hat{ {\up}}_x\Big\}  \\
    \label{eq:gy_derive}
    \hat{\Gamma}_y &=&  \hat{\Gamma}'_y + \hat{\Gamma}''_y
    = \frac{1}{2i\hbar}
    \Big\{\hat{\Theta}_R,\hat{ {\up}}_y\Big\} + \frac{1}{Ri\hbar}\Big(\hat{r}_x\hat{p}_y-\hat{r}_y\hat{p}_x-\frac{R}{2} \Big\{\hat{\Theta}_R, \hat{p}_y\Big
\}\Big)= \frac{\hat{l}_z}{Ri\hbar}\\
\label{eq:gz_derive}
    \hat{\Gamma}_z &=&  \hat{\Gamma}'_z + \hat{\Gamma}''_z
    = \frac{1}{2i\hbar}
    \Big\{\hat{\Theta}_R,\hat{ {\up}}_z\Big\} - \frac{1}{Ri\hbar}\Big(\hat{r}_z\hat{p}_x-\hat{r}_x\hat{p}_z +\frac{R}{2} \Big\{ \hat{\Theta}_R,\hat{p}_z\Big\}\Big) = -\frac{\hat{l}_y}{Ri\hbar}\label{eq:Gammas_final}
\end{eqnarray}
\end{widetext}
Here,  we define $\hat{\Theta}_R$ as
\begin{eqnarray}
    \hat{\Theta}_R &=& \frac{M_2\hat{\Theta}_1-M_1\hat{\Theta}_2}{M_1+M_2}\label{eq:thetaR}
\end{eqnarray}
which transforms just like $\hbG_R$.
For a derivation of Eqs. \ref{eq:gy_derive}-\ref{eq:gz_derive}, see Appendix \ref{appendix:erf_cals}.
In the end, 
Eqs. \ref{eq:compare:bunker}-\ref{eq:Gammas_final} are the final equations that are implemented below and compared against exact calculations.

Note that Eq. \ref{eq:compare:bunker} above has some resemblance to an exact reduced expression for a diatomic molecule\cite{Bunker2006}, but that resemblance is far from exact. That being said, Eq. \ref{eq:gamma_general} can easily be extended and applied to polyatomic and even material systems (which is more exciting). Interestingly, note also that $\hat{\Gamma}_y$ and $\hat{\Gamma}_z$ have no dependence on $\hat\Theta_R$; they depend only on the total electronic angular momentum.  In fact, the form of these equations relates strongly to the famous diatomic work of Moody, Shapere and Wilczek\cite{Moody1986}, which we discuss in Sec. \ref{sec:MSW} below.

\subsubsection{Vibrational energies and Nuclear/Electronic Observables}
In this article, our goal is to compare the nuclear+electronic eigenenergies and eigenvectors of the exact hamiltonian vs those from BO and/or phase space electronic strucure approximations. Now, in order for a phase space calculation to predict a quantized nuclear+electronic  wavefunction, one must requantize the classical nuclei. To that end, if we are working on the electronic ground state $\hat{H}_{\PS}$ with energy
$E_{\PS}^0$ and wavefunction $\Psi_{\PS}^0$, we will requantize the nuclear degrees of freedom and generate a quantum vibrational Hamiltonian by applying a Weyl transform:  
\begin{equation}
    \bra{R}\hat{H}_{\PS}^{\vib}\ket{R'} = \int \frac{dP_R}{2\pi\hbar} e^{\frac{i}{\hbar}P_R\cdot(R-R')} E_{\PS}^0\left(\frac{R+ R'}{2},P_R\right)\label{eq:Weyl}
\end{equation}
Note that Eq. \ref{eq:Weyl} is an integral over only one dimension. In principle, one might like to Weyl transform over both the two nuclear rotational angles and the vibration, but in practice a Weyl transform works only for motion in a cartesian frame. 
Moreover, we have already discussed above how to choose $(P_\vartheta,P_\phi)$ so as to map PS energies and eigenstate to their corresponding exact analogues as a function of $J$. The $j^{th}$ vibrational eigenfunction of the (zeroth) ground electronic state $\hH^{\vib}_{\PS}$ is defined by:
\begin{eqnarray}
    \hat{H}^{\vib}_{\PS} \ket{\chi_j^0} = E_j^0\ket{\chi_j^0}
\end{eqnarray}

Below, we will report not only eigenstate energies but also various electronic and nuclear observables. To that end, after performing the Weyl transform and diagonalizing $\hat{H}_{\PS}^{\vib}$, nuclear expectation values are calculated directly as in any other quantum mechanical calculation.  Electronic expectation values for an operator $\hat{O}_e$ are  calculated by  $(i)$
evaluating the expectation value of the Wigner transform $\hat{O}_e$  at every point in nuclear phase space, $(R,P_R)$,  $(ii)$ applying a Weyl transform to convert the resulting  phase space distribution back to position coordinates, and $(iii)$ averaging over the vibrational quantum eigenstates:

\begin{widetext}
 
\begin{eqnarray}
\hat{O}_{e}^{W}(R,P_R) &= \bra{ \Psi_{\PS}^0(R,P_R)} \hat{O}_e \ket{\Psi_{\PS}^0(R,P_R)}\\
\bra{R}\hat{O}_{e}\ket{R'} &= \int \frac{dP_R}{2\pi\hbar} e^{\frac{i}{\hbar}P_R(R-R')} \hat{O}_{e}^{W}\left(\frac{R+R'}{2},P_R\right) \\
\langle\hat{O}_{e}\rangle_{01} &= \int dR dR' \braket{\chi_0}{R}\bra{R} \hat{O}_e\ket{R'}\braket{R'}{\chi_1}
\end{eqnarray}
   
\end{widetext}

As with the exact calculation, additional details about the implementation of the PS calculation are included in Appendix \ref{appendix:numerics}.

\section{Born-Oppenheimer Revisited \label{sec:BO}}
In what follows below, we will compare eigenvalues and observables for phase space electronic structure calculations vs. exact calculations. We will also compare against BO calculations on one electronic surface (no Born-Huang expansion) as is practical for most quantum chemistry simulations of realistic systems, but here there is an important nuance. In a naive fashion, one could construct a BO Hamiltonian that ignored all rotational motion:
    \begin{eqnarray}
        \hH_{\BO}(P_R,R)  &=& \frac{1}{2\mu_{R}}P_R^2    + \hat H_{\el}(\hat x,\hat y,\hat z) 
    \end{eqnarray}
But of course, this naive ansatz ignores all kinetic energy in the rotational motion. A better BO approximation is to set:
\begin{align}
        \hH_{\BO}^J(P_R,R)  &= \frac{1}{2\mu_{R}}\Big(P_R^2 + \frac{\hbar^2 J(J+1)}{R^2}\Big)  \nonumber \\&\hspace{1cm}\nadine{+ \hat H_{\el}(\hat x,\hat y,\hat z) }
    \label{eq:BOJ}
\end{align}
In section \ref{sec:results}, we will compare against Eq. \ref{eq:BOJ} (under the label `BO') below.

Within the BO approximation, it is common to identify a diagonal BO correction (DBOC), which is formally $(H^{\BO}_{\DBOC})_{00} =\frac{1}{2\mu_R}\sum_k d^R_{0k}\cdot d^R_{k0}$ (written out in terms of a sum over all adiabatic electronic states). Below, we will estimate this correction by working only with derivatives along the the internuclear axis:
\begin{align}
H^{\BO}_{\DBOC} \approx \frac{-\hbar^2}{2\mu_R} \left<\Psi^{\BO}_0(\uvr;R) \middle| \partial_{R}^2 \Psi^{\BO}_0(\uvr;R) \right> \label{eq:DBOC}
\end{align}
In the results section below, we will sometimes add the term above to Eq. \ref{eq:BOJ} to test whether the DBOC  can provide corrections to vibrational energies and nuclear observables. We note that, as written, this approximation of the DBOC cannot directly affect electronic observables (such as an electronic centrifugal energy).

\section{Results}\label{sec:results}
\begin{table*}
    \centering
\begin{tabular}{|c|c|c|c|c|c|c|c|}
\hline
    D & d & a & A & B & C & $M_1$ & $m_e$\\
    \hline
    60 kcal/mol & 0.95 {\AA} & 2.52 {\AA}$^{-1}$& $1.16\times 10^5$ kcal/mol & 3.15 {\AA}$^{-1}$ & $1.155\times 10^4$ kcal/mol/{\AA}$^{-6}$ & $1\times 10^6$ amu & 1 amu \\
    \hline
\end{tabular}
    \caption{Parameters of the potential and reference mass for sample calculations. Adapted for higher dimensions from Ref \cite{Borgis2006}.}
    \label{tab:potential}
\end{table*}

\begin{figure*}
\centering\includegraphics[width=\linewidth]{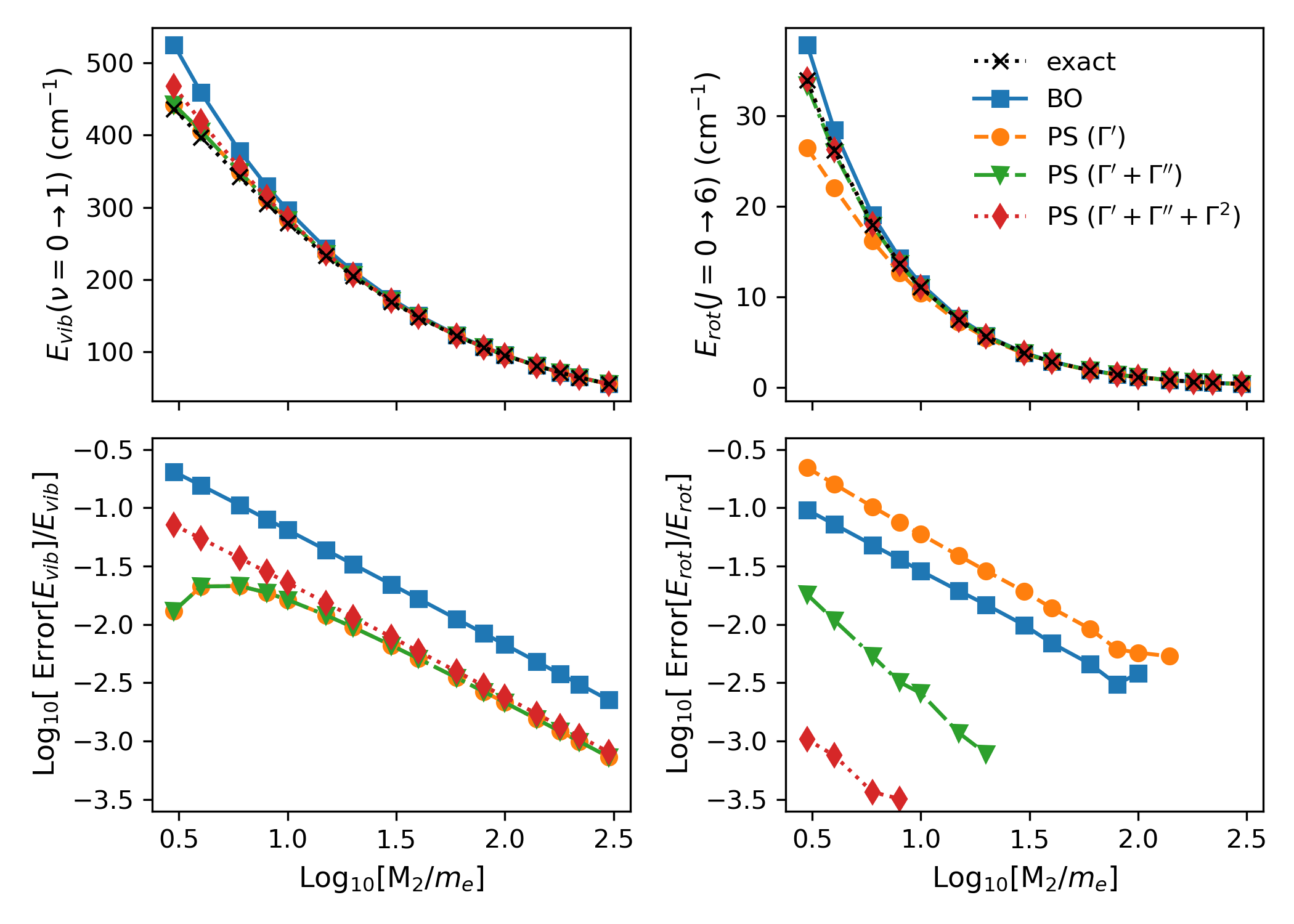}
    \caption{Vibrational (top left)  and  rotational energy (top right) transitions  for the model system as plotted as a  function of  log of the mass ratio. The data shown here is for fixed rotational state $J=0$ for all calculations. PS vibrational calculations are from Eq. \ref{eq:HPS_final} and BO calculations are from Eq. \ref{eq:BOJ}.  For rotations, the gaps are already numerically very small, so in-order to maximize the resolution of our Davidson solver, we compare ground state energies at fixed $J=6$ vs $J=0$, even though this would not be a a quantum mechanically allowed transition. In the bottom row, we plot the log of the relative error (relative to the exact vibrational and rotational gaps) as a function of the mass ratio. Note the strong performance of a PS electronic structure approach.}
    \label{fig:energy_gaps}
\end{figure*}

We have implemented the algorithms above as described. In order to 
benchmark phase space electronic structure across a wide array of potentials (with varying degrees of non-adiabaticity),  we have selected to use a potential previously designed for the study hydrogen bonding \cite{Borgis2006, Scherrer2017}. 
To access a wide array of adiabatic or nonadiabatic regimes, we made one of the two nuclear masses  exceedingly large, $M_1 = 10^6$ atomic mass units (amu), while the mass of $M_2$ is varied multiple orders of magnitude and can be even as light as the light particle or 'electron' at 1 amu.
Use of such a large $M_1$ guarantees the traditional chemical ordering of transitions is preserved even when $M_2 = m_e$: namely, that rotational transitions are smaller in energy than vibrations, which are subsequently smaller than electronic transitions. 
Our final form is:
\begin{align} 
 V(\uvR,\uvr) &= D\left(e^{-2a(|\uvr +\frac{\mu_R}{M_2}\uvR|-d)} - 2 e^{-a(|\uvr+\frac{\mu_R}{M_2}
\uvR|-d)}\right) \nonumber \\
 &+ \frac{D}{2}\left(e^{-2a(|\uvr-\frac{\mu_R}{M_1}\uvR|-d)} - 2 e^{-a(|\uvr-\frac{\mu_R}{M_1}\uvR|-d)}\right) \nonumber \\
    &+ Ae^{-B|\uvR|} - \frac{ C}{|\uvR|^6} 
\end{align}
The parameters for this potential and the reference mass are given in Table \ref{tab:potential}.

All results in this section are for the three-dimensional calculation; results of our calculations in two dimensions are given in Appendix~\ref{appendix:2D_data}.

\subsection{Vibrational Energies}

On the upper left hand side of Figure \ref{fig:energy_gaps},  we report the vibrational energy gaps of the model system across multiple orders of magnitude of the mass ratio between $M_2$ and $m_e$. For every system, the first excited vibration in the system at a fixed $J$ is the first overall excited state of the system. 
On the bottom left hand side of Figure \ref{fig:energy_gaps}, 
we plot the log of the relative error (relative  to the exact calculation). For  the vibrational energy, we find that both BO and PS models are reasonably accurate, but show a logarithmic slope of $-1$ because the missing components from these theories both scale with the inverse mass ratio, $(m_e/M_2)$. PS consistently performs a whole order of magnitude better than BO, indicating that we have indeed captured some of the electron nuclear coupling. 

\subsection{Rotational Energies}

Next, on the upper right hand side of Figure \ref{fig:energy_gaps}, we plot the rotational energy gap.  Because all calculations are performed at a fixed $J$, here we must compare ground state energies for different fixed $J$
in order to extract a rotational energy gap.  We plot the relative error of the energy gap in the lower right hand side of Figure \ref{fig:energy_gaps}. 
Here, we find that  the inclusion of the electron rotation factor ($\hat{\bm \Gamma}''$) leads to a major  increase in accuracy.  
With the addition of the $\hat{\bm{\Gamma}}^2$ term, we can detect the error in our rotational transitions only for the lightest mass ratios; otherwise,  for all larger mass ratios, the errors are in fact below the precision of our Davidson solver at ($10^{-12}$ au $= 2.2 \times 10^{-7} \;\mathrm{cm}^{-1}$. Such powerful accuracy for rotations indicate the that the $\hat{\bm \Gamma}''$ and $\hat{\bm \Gamma}''\,^2$ corrections really do capture the Coriolis $({\hat{\bm{J}}}\cdot{\hat{\bm{l}}}_r)/(\mu_R R^2)$  and centrifugal $({\hat{\bm{l}}}_r\cdot{\hat{\bm{l}}}_r)/(\mu_RR^2)$ coupling that appear in the exact Hamiltonian. Additional data for the 2-dimensional calculation, which shows strikingly similar trends, can be found in Appendix \ref{appendix:2D_data}

\begin{figure*}
    \centering
    \includegraphics[width=\linewidth]{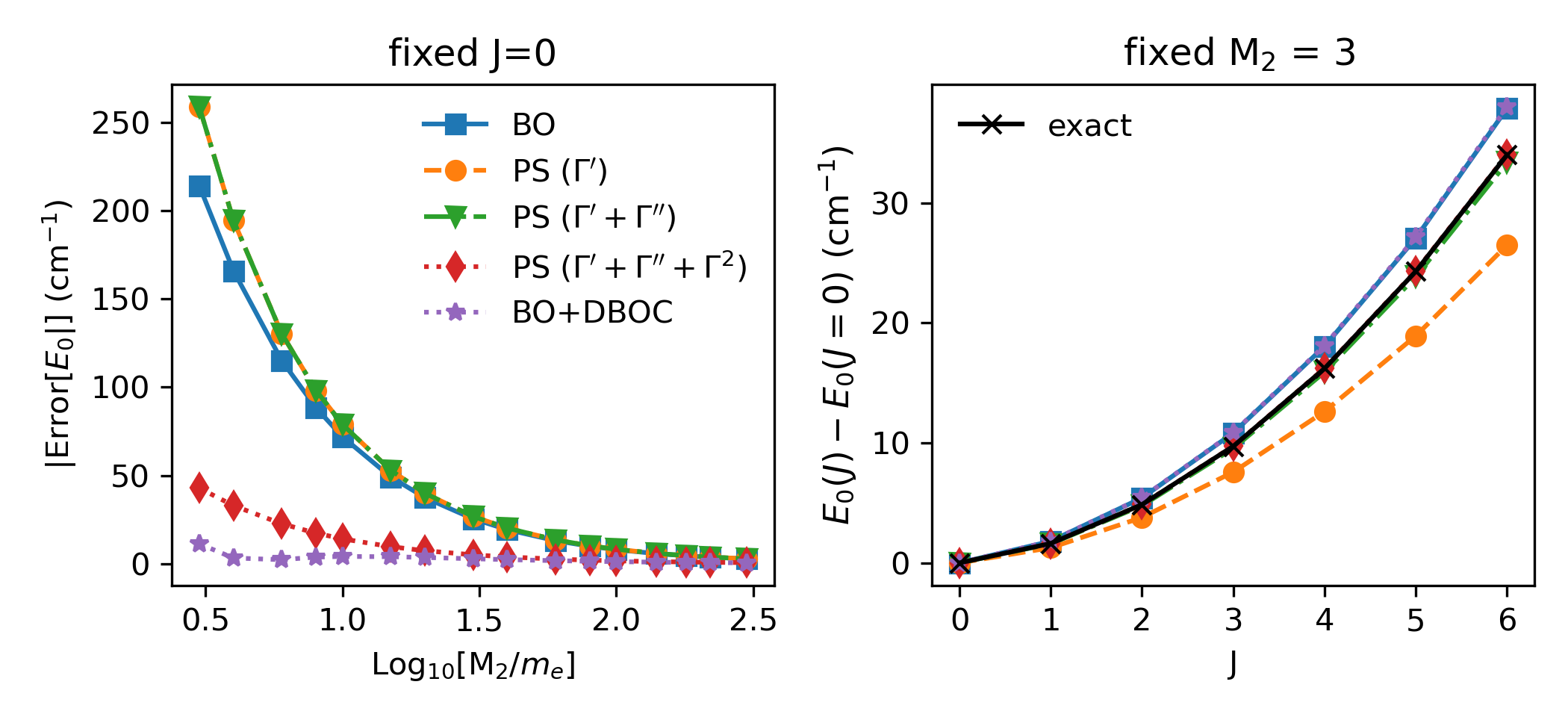}
    \caption{Error in the ground state energy as a function of mass ratio (left) and ground state energy as total angular momentum $J$ (right). }
    \label{fig:dboc}
\end{figure*}

\subsection{The Second Order PS Term and Exact Ground-State Energies}

The data on the bottom right of Fig. \ref{fig:energy_gaps} indicates that the 
 $\hat{\bm \Gamma}^2$ term can have a large effect on a calculation. 
 In our simulations (see  left hand side of Figure \ref{fig:dboc}), we find that the effect of including the DBOC (Eq. \ref{eq:DBOC}) is to reduce the absolute error of the ground state energy of the system at $J=0$.  On the right hand side of Fig. \ref{fig:dboc}, however, we further plot  the ground state  energy as a function of $J$
for the most non-adiabatic mass ratio of $M_2/m_e = 3$, and we find that the effect of the DBOC is marginal. Thus, it would appear that inclusion of the DBOC apparently contributes only to fixing up absolute (as opposed to relative) energies.  Turning to a PS approach, encouragingly, we find that including the  $\hat{\bm \Gamma}^2$ term can incorporate these DBOC attributes. Perhaps most importantly, even though  inclusion of the $\hat{\bm \Gamma}^2$ term alone cannot solve the  BO electronic momentum problem, we will show below that 
including the  $\hat{\bm \Gamma}^2$ term (in tandem with the $\hbP \cdot \hbG$ terms) can improve many other  observables.

\begin{figure*}
        \centering
    \includegraphics[width=\linewidth]{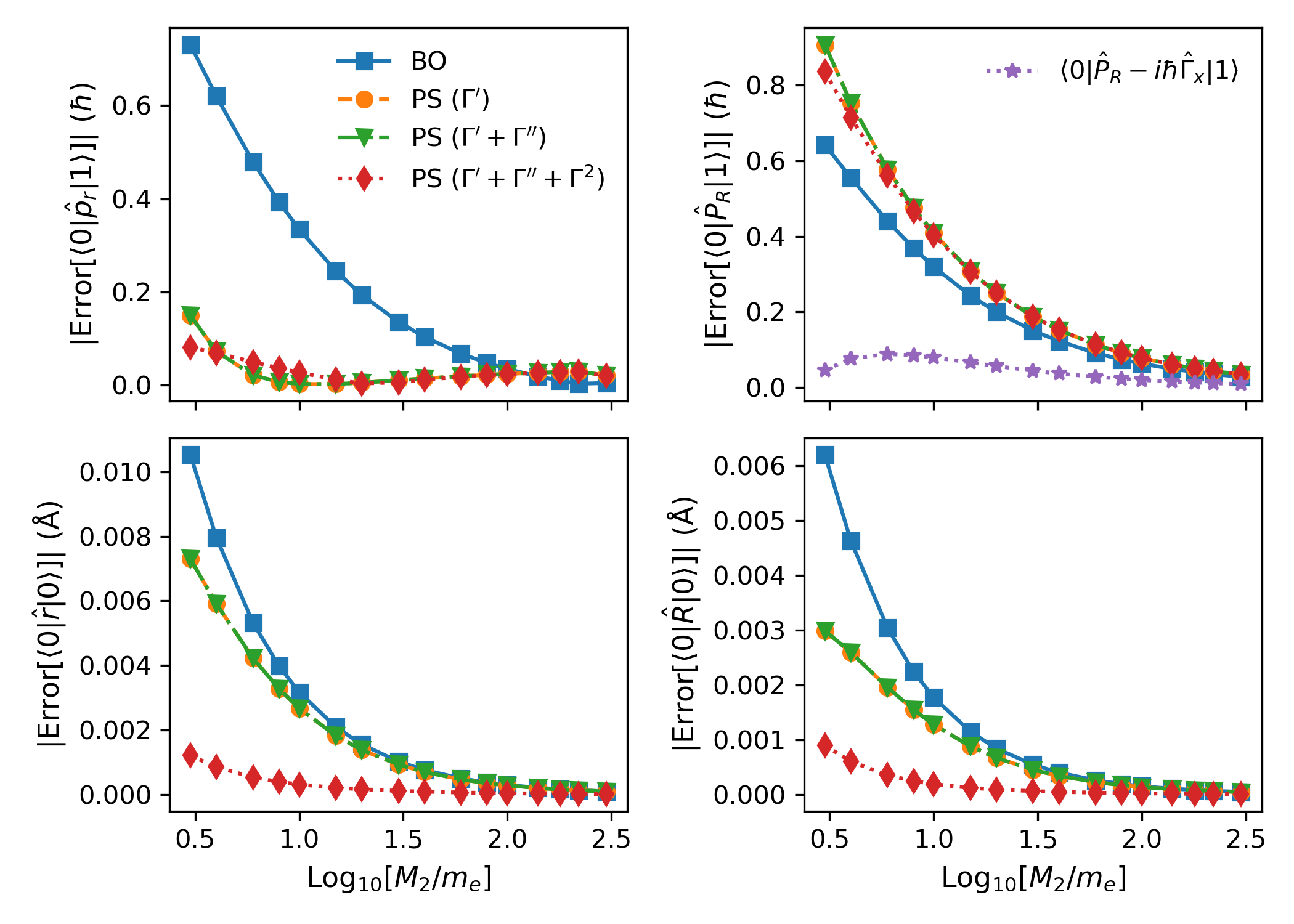}
    \caption{Data for linear observables. Absolute errors vs mass ratio for matrix elements of the electronic and nuclear linear momentum and position operators at $J=6$. For the case of the electronic linear momentum between the ground and first vibrational state, $\langle 1 | \hat{p}_r|0\rangle$, we calculate this expectation along the radial direction of the electron. In BO theory, this electronic momentum is completely missing, so the magnitude of the error represents the exact electronic momentum.   Following the analysis in Sec. \ref{sec:linearmomentum}, PS theory  maps the exact nuclear momentum to $\left<\hbP - i\hbar\hbG\right>$ rather than just $\left< \hbP \right>$ alone, as found previously in Ref. \cite{Bian2025-vib}.
    For the electronic and nuclear position (bond length), we find that all PS theories show a marked improvement over BO theory.}
    \label{fig:vib_errors}
\end{figure*}

\begin{figure*}
    \centering
    \includegraphics[width=\linewidth]{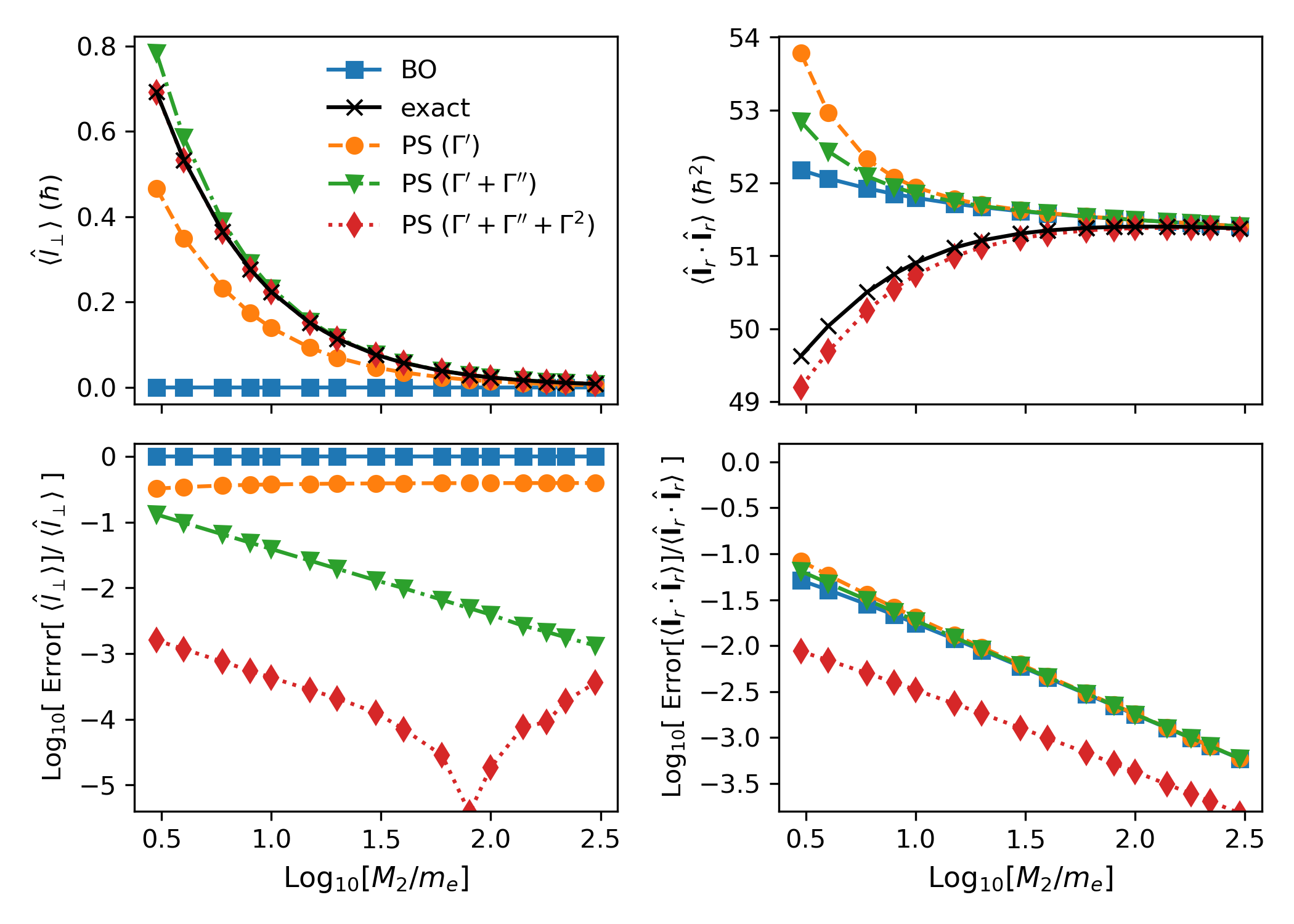}
    \caption{Electronic angular momentum in the ground state for different kinds of phase space calculations vs the log of the mass ratio for a fixed $J=6$ in all plots. We plot both the linear $\langle \hat{l}_\perp\rangle$ along one of the axes perpendicular to the nuclear axis (which is completely missing from BO theory) as well as $\langle \hat{\bm l}^2\rangle$ (which is a real valued operator and exists in this potential in the ground state even for $J=0$). In the bottom row, we plot the log of the relative error (relative to  the exact angular momenta) for $\langle\hat{l}_\perp\rangle$ and $\langle \hat{\bm l}^2\rangle.$
    Note the very strong performance of PS electronic structure theory.}
    \label{fig:ang_mass}
\end{figure*}

\begin{figure*}
    \centering
    \includegraphics[width=\linewidth]{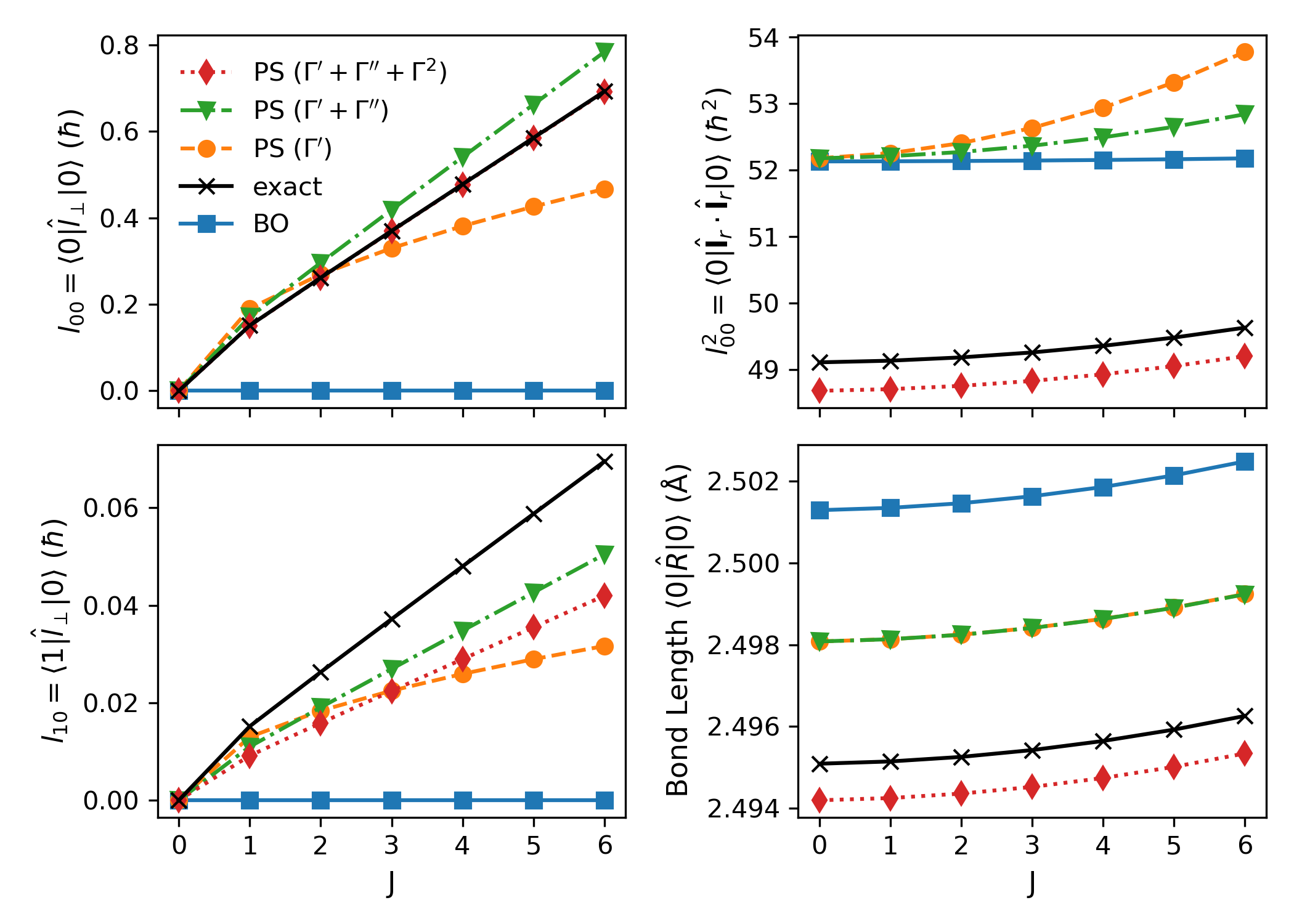}
    \caption{Electronic angular momentum data for $M_2/m_e = 3$ as a function of changing J. We show that the full PS theory with $(\hbG' + \hbG'' + {\hbG}^2)$ corrections is able to fully capture the change in angular momentum for $\langle\hat{l}_\perp\rangle$ and $\langle\hat{l}^2\rangle$ in the ground state. For a transition angular momentum between vibrations $\langle1 |\hat{l}_\perp|0\rangle$, PS is only partially able to recover the change in angular momentum. Lastly we show the bond length of the nuclei in the ground states as a function of $J$. We find that all theories including BO are correctly able to capture the expansion of the bond with increasing $J$, but overall these are small effects with only slightly different bond length at $J=0$. }
    \label{fig:ang_J}
\end{figure*}

\subsection{Linear Observables: Linear Momentum and Position}\label{sec:linearmomentum}

Before we address the most interesting (angular) observables, let us address the linear observables that have previously been studied in one-dimensional models \cite{Bian2025-vib}
In Fig. \ref{fig:vib_errors}, we plot (i)  the linear electronic momentum along the distance from the nuclear center of mass (the radial component of $\hat{\bm p}_e$),  (ii) the linear nuclear momentum  along the internuclear direction (iii) the radial component of the electronic position and (iv) the nuclear bond length.  We  show that even in a rotating system with $J=6$, the $\langle 1 | \hat{p}_r |0\rangle$ and $\langle 1 |\hat{r} | 0\rangle$ matrix elements are always well captured by  PS electronic structure theory---implying that the terms $(\hbG'')$ do not influence linear motion along the nuclear axis (vibration). Note that whereas PS theory recovers the electronic momentum between vibrational states of the system, $\langle 1 |\hat{p}_r|0\rangle$, this quantity vanishes within BO theory. 

In the case of the nuclear momentum, as was predicted earlier \cite{Bian2025-vib}, we find that  one {also} recovers an accurate answer only by calculating $\langle\uvP_R - i\hbar \hbG'\rangle$. In other words, within a PS model, the nuclear momentum represents the total linear momentum of the molecule, $\uvP_R^{\PS} \approx \uvP_R + \hat{\uvp}_e$, just as shown for BO adiabatic theory in Ref. \cite{Littlejohn2024,yanzewu:2024:jcp:pssh_conserve}.
This finding is the linear momentum analogue to the angular momentum interpretation that was used to construct the notion of a PS ``body-frame'' in the above calculations (see Sec. \ref{sec:PS_Ptheta}).

\subsection{Angular Momentum Observables}

Finally, we turn now to the most exciting and entirely new calculations, namely the case of angular observables (for which  one-dimensional analogues are impossible). In Figure \ref{fig:ang_mass}, we plot the expectation value of $\hat{l}_{\perp}$ and $\hat{\bm l}^2$  for different mass ratios of the nuclei for a $J=6$ state. Physically, the meaning of $\langle \hat{l}_\perp\rangle$ is clear for a diatomic, but mathematically the form depends on the different gauge choices for the exact and PS calculations. For the exact calculation, $\hat{l}_\perp = \hat{l}_a$. This definition follows because the body frame coordinate exact wavefunction $\xi^J_{\ell\Omega}(R,r)$ is real valued due to Eq. \ref{eq:3D:Hxi_diag} being real valued, and thus  $\hat{l}_b = -\frac{i}{2}(\hat{l}_- - \hat{l}_-)$ is a purely imaginary operator in the spherical harmonic basis satisfying $\langle \hat{l}_b\rangle = 0$. For the PS calculation $\hat{l}_\perp$ is chosen to be $\hat{l}_y$ due to the choice that $P_\phi=0$.
From the data in Fig. \ref{fig:ang_mass}, we find that PS  performs quite well here. Note that the electron translation factors ($\hbG'$) do capture some of the angular momentum of the system ground state, but 
clearly the electron rotation factors ($\hbG''$) are essential for accurate electronic angular momenta
at all mass ratios. Most importantly, notice how different are the slopes of the different curves  in the bottom left of Fig. \ref{fig:ang_mass}. 
The same conclusion holds if we look at the data as a function of different  $J$ values on the top left hand of Fig. \ref{fig:ang_J}.

Next, we consider $\langle { \hat{\bm l}}_r \cdot { \hat{ \bm l}}_r\rangle$. This operator is a real operator and thus can be non-zero within BO theory.
Data along the electronic ground state are plotted on the right of Fig. \ref{fig:ang_mass}. Here, especially from the log data, it is clear that a PS theory does not capture accurately capture the $\langle { \hat{\bm l}}_r \cdot {\hat{\bm l}}_r\rangle$ without the $\hat{\bm{\Gamma}}^2$ term. The latter 
offers a correction that scales as  $(m_e/M_2)$ (just as for the $\hbP \cdot \hbG$ term), coming from the centrifugal correction to the wavefunctions $\propto {\bf \hat{l}}_r \cdot {\bf \hat{l}}_r/2\mu_RR^2$ as seen in the exact body frame coordinate Hamiltonian.  This same conclusion also can be found when look at $\langle {\bf \hat{l}}_r \cdot {\bf \hat{l}}_r\rangle$ as a function of $J$  in Fig. \ref{fig:ang_J}. 

Lastly, in Fig. \ref{fig:ang_J}, beyond observables already mentioned, we also plot the off-diagonal matrix element of 
the angular momentum perpendicular to the internuclear axis of the molecule $\langle\hat{l}_\perp\rangle_{01}$ and the nuclear bond length $\langle R\rangle$
as a function  of increasing $J$ (keeping mass fixed).  For the former, the data is still good (though increasing the second order term seems to hurt just  a bit); for the latter, the second order corrections helps quite a lot (though the changes do seem small in absolute magnitude).

Overall, the data from Figs. \ref{fig:energy_gaps}-\ref{fig:ang_J} is striking: with the correct choice of $\hbG''$, PS electronic structure theory can achieve the same level of accuracy as was found previously for one-dimensional models \cite{Bian2025-vib} but with much richer physical observables possible.

\section{Discussion: Connection to Moody-Shapere-Wilczek \label{sec:MSW}}

Thus far, we have not discussed the non-abelian Berry curvature of the $\hbG$ operators. As a reminder, when working with derivative couplings, the non-abelian Berry curvature $\Omega$ (not to be confused with $\Omega$ the $\hat J_c$ eigenvalue in the exact calculations) is defined as\cite{mead:1992:rmp}
\begin{align}
\Omega_{jk}^{A\alpha,B\beta} = (\partial_{X_{A,\alpha}} d^{B\beta}_{jk} - \partial_{X_{B,\beta}} d^{A,\alpha}_{jk}) + \Big[d^{A,\alpha}, d^{B\beta}\Big]_{jk}
\end{align}
where the added index $\alpha,\beta \in \{\hat x,\hat y,\hat z\}$ for nuclei $A,B$.
The $\Omega$ transforms as a tensor and characterizes a subspace of electronic states. In the limit of a complete basis, it follows that $\Omega = 0$; but in an incomplete basis, $\Omega$ is finite.

In the specific context of a diatomic molecule, Moody, Shapere and Wilczek (MSW) \cite{Moody1986} showed many years ago that the non-abelian curvature did not vanish for a rotational multiplet with electronic angular momentum. In fact, let
\begin{align}
        F^\gamma_{jk} \equiv \frac{1}{2}\sum_{\alpha \beta} \epsilon_{\alpha \beta \gamma}
\Omega_{jk}^{A\alpha,B\beta}
\end{align}
If one freezes the nuclear bond length and calculates the curl along the nuclear axis $\hat x$, the result is that, for a doubly degenerate electronic system with electronic angular momentum $l \ne \frac{1}{2}$, 
$F^x$ is a magnetic monopole in the relevant two-dimensional subspace:
\begin{align}
\left( \hat{F}_x^{MSW} \right)_{2 \times 2} &= -\left( \frac{i\,\hat{l}_x}{R^2\hbar} \right)_{2 \times 2} \label{eq:MSW}
\end{align}

With this background in mind, one can ask:
What do we expect from the corresponding curvature of the $\hbG$ operators? How well do the $\hbG$ operators mimic the derivative couplings? 
 The non-abelian curl of $\hat \Gamma$  is defined as \footnote{Note curl terms involving $\hbG^{\NCM}$ all vanish so we only have to consider terms with $\hbG^R$ in them.} 
\begin{align}
\hat{F}_x = \hat{\Omega}^{R y,R z} \equiv \frac{\partial \hat{\Gamma}_{z}}{\partial R_{y}} - \frac{\partial \hat{\Gamma}_{y}}{\partial R_{z}} + \left[\hat{\Gamma}_{y}, \hat{\Gamma}_{z}\right] 
\label{eq:def_jodd}
\end{align}
and cyclically thereafter. Note that Eq. \ref{eq:def_jodd} is an electronic operator, defined in the entire Hilbert space (and not a 2-dimensional subspace like Eq. \ref{eq:MSW}). That being said, the meaning of the curl is the same in both contexts---it acts like a pseudomagnetic field on the relevant nuclear degrees of freedom. See Appendix \ref{sec:appendix_curl}.

Now, starting from the definitions in Eq. \ref{eq:gx_derive}-\ref{eq:gz_derive}, a simple calculation shows that:
\begin{align}
    \hat{F}_x &= -\frac{i\,\hat{l}_x}{R^2\hbar}\label{eq:Fx}\\
\hat{F}_y &= -\frac{i}{2 R \hbar}\left\{
\hat z\frac{\partial \hat{\Theta}_R}{\partial x},{\hat p_x}\right\}-\frac{i\hat{\Theta}_R}{R \hbar }\hat{p}_z+\frac{i\hat{l}_y}{R^2\hbar}\\
\hat{F}_z &= +\frac{i}{2 R \hbar}\left\{ \hat y\frac{\partial\hat{\Theta}_R}{\partial x} ,{\hat p_x}\right\}+\frac{i\hat{\Theta}_R}{R \hbar }\hat{p}_y+\frac{i\hat{l}_z}{R^2\hbar}\label{eq:Fz}
\end{align}

Two conclusions follow from the calculation above. First, let us ignore all the terms 
proportional to $\hat{\Theta}$.  In this case, we find that 
our proposed phase space electronic structure theory equations of motion  are in exact agreement  with MSW, insofar as they reduce to the exact same matrices over the relevant two-dimensional subspace.  In particular,  $\hat{F}_x \equiv \hat{\Omega}^{R y,R z}$  represents a magnetic monopole present. Furthermore, if we look at $\hat{F}_y$ and $\hat{F}_z$, we will find that these terms vanish because $\left( \hat{l}_y \right)_{2 \times 2} =\left( \hat{l}_z \right)_{2 \times 2} = 0$ for a system with $l \ne 1/2.$ Thus, our phase space electronic structure theory agrees exactly with  MSW on the relevant electronic subspace -- but interestingly, the operator is defined on a far larger hilbert space.  As a side note, we mention that  for the most part, the electronic states investigated  in this paper 
have only a small amount of electronic angular momentum, and so further numerical studies and benchmarking will still be fruitful.

Second, let us address the terms proportional to $\hat{\Theta}$.  These terms arise from the $\hat{\Theta}$ functions in Eq. \ref{eq:gx_derive} and electronic motion coupled to the internuclear vibration (ro-vibronic coupling). These  terms  are  not relevant to the MSW curl (within a two-dimensional degenerate subspace), but are critical to recover the vibrational results in Fig. \ref{fig:energy_gaps}. 
Interestingly, note that if we set $\Theta_R = \hat{x}/R$ (which is very similar to the  dilation factor used in Ref. \cite{polkovnikov:2026:pnas} for a particle in a box), then we find that 
$\hat{F}_y = \hat{F}_z = 0$.
In other words, establishing a meaningful phase space electronic structure theory capable of treating vibrations requires going beyond a linear switching function and establishing $\hat{\theta}$ functions with firmer boundaries between different atomic domains.

In the end, we submit that Eqs. \ref{eq:Fx}- \ref{eq:Fz} above are a rigorous means to include many aspects of the derivative coupling directly into an electronic structure calculation for a rotating {\em and vibrating} diatomic.

\section{Conclusion}\label{sec:conclusion}

In this work, we have investigated a diatomic molecule through both phase space electronic structure theory and BO theory. In particular, we have derived a new, phase space electronic Hamiltonian for a diatomic (Eq. \ref{eq:compare:bunker}) that goes well-beyond Born-Oppenheimer theory and allows one to solve for a basis of electronic eigenstates as a function of the total, molecular angular momentum.  In order to better understand the nature of the resulting eigenstates in the presence of low-lying, or potentially degenerate, electronic states, we have evaluated eigenvalues and other observables (especially electronic observables) in the presence of artificially low nuclear/electronic mass ratios---a regime where electronic excitation energies need not be much larger than vibrational electronic energies. We have further gone far beyond the previous work in Ref. \cite{Bian2025-vib} investigating models of hydrogen transfer, as we now work in three dimensions (rather than one dimension).  Noninertial frames are far more interesting (and difficult to account for) in three dimensions---where coriolis forces and centrifugal forces appear, and where developing exact benchmark calculations is far more involved. Here, we have developed an exact code for the three-body problem following Ref. \cite{Bian2025-vib} and boosted the resulting algorithm for performance by implementation on an A100 80 GB GPU. The full source code is available at \cite{shabica_2026_20047795} and can be freely used for further benchmarking by other chemists and physicists.

The data presented above is the first definitive test of a multidimensional $\hbG$ operator\cite{Tao2025_basis_free}, as written down explicitly for a diatomic in three dimensions in Sec. \ref{sec:etferf}; one cannot fully benchmark $\hbG''$  and Coriolis forces without exact 3D calculations.   From the strength of the data presented above, we can unambiguously conclude that  phase space electronic structure theory can indeed be applied to   ro-vibrational transitions and the resulting algorithm provides substantially improved energies and wavefunctions relative than traditional BO theory. This finding is especially important because spin degrees of freedom are the most common source of electronic degeneracy and modern research points out that angular momentum exchange, especially involving spin degrees of freedom,  manifests itself in very interesting, a la mode experiments\cite{naaman:2015:arpc}. Thus, the fact that phase space electronic structure theory has been shown to predict accurate angular momenta  here (just as  Ref. \cite{Bian2025-vib} achieved working with linear momenta) is a strong endorsement of incorporating phase space electronic structure theory into modern electronic structure packages.  


Looking forward, obviously the next practical step is indeed to implement  phase space electronic structure theory into efficient {\em ab initio} software packages -- either using atomic orbitals or plane waves. 
One big conceptual and practical question will be the implications of the $\hbG \cdot \hbG$ term identified above in Sec. \ref{sec:gammasq}, that introduces a new two-body electron-electron interaction into the electronic Hamiltonian. One must wonder whether this interaction leads to superconductivity in some fashion\cite{tinkham_superconductivity}.

Another big conceptual question will be to extend the present research to systems with explicit spin degrees of freedom. As noted in the introduction, here we have been able to benchmark $\hbG'$ and $\hbG''$, but in the presence of spin, the $\hbG$ operator should also include a third term $\hbG'''$ which ensures that the total conserved angular momentum includes spin. Physically, the inclusions of $\hbG'''$ is equivalent to making sure that when one walks along a given adiabatic, the frame for the spin direction rotates with the molecule -- rather than staying fixed in the lab frame.  Within the small-molecule community, $\hbG'''$ is known as the S-uncoupling operator.\cite{field2015spectra} Physically, $\hbG'''$ ensures the presence of a spin-coriolis that has been imputed for many exciting spin effects, especially the 
chiral induced spin selectivity (CISS)
effect.  As a reminder, the CISS effect emerges when electrons are conducted across chiral media and show a spin preference for the transport, and there is now the thought that chiral phonons and angular momentum exchange between electrons and nuclei is essential. To that end, future extensions of phase space electronic structure theory to include spin will be very helpful.

\section*{Acknowledgments}

Computing support for this came from Princeton Research Computing and the 2025 Princeton Open Hackathon. 
We have also used resources
of the National Energy Research Scientific Computing Center
(NERSC), a U.S. Department of Energy Office of Science User
Facility operated under Contract no. DE-AC02-05CH11231.
We would like to additionally thank Cameron Khan, Xinchun Wu, and Alok Kumar for their contributions to the code development and hackathon participation.
This work was
supported by the U.S. Department of Energy, Office of Science, Office of Basic Energy Sciences, under Award No. DE-SC0025393.

\appendix
\onecolumngrid
\vspace{\columnsep}
\section{Details of the numerical solutions\label{appendix:numerics}}

The code for this project is available at Ref \cite{shabica_2026_20047795}. 

\paragraph{Exact calculation details:}

Calculations presented for 2D are done on grids for the three coordinates $(R,r,\gamma)$, for a grid size of $N_R = 90$, $N_r = 100$, $N_\gamma = 90$. We find the number of grid points is adequate to converge all mass and $J$ points considered. While both $R$ and $r$ are defined as radial grids, in practice, the bond length $R$ never goes to 0, additionally for this potential, we find $r_{max} = R_{max}$ is acceptable for convergence. The grid extents for $R$ coordinate was determined at $J=0$ and a set of reduced masses given in Table \ref{tab:extents}, and linearly interpolated intermediate masses.

The kinetic energy for $R$ coordinate was represented with a spectral kinetic energy DVR, as this formalism matches best with the nuclear kinetic energy formed by a Weyl transform of $\bP_R^2$, as given in Eq. \ref{eq:Weyl}. For the $r$ coordinate, a radial coordinate Colbert-Miller\cite{Colbert1992} sine-function DVR for the kinetic energy was chosen. While this DVR is not ideal when there is {considerable} electronic density at $r\to0$, this is not a problem in asymmetric mass regimes where $M_1 \neq M_2$. As done in \cite{Kuppermann1976_2D}, the exact wavefunction $\zeta_J$  is typically re-scaled to remove the Jacobian on the radial coordinate in 2D,

\begin{align}
    \zeta_J(R,r,\gamma) &= \frac{1}{\sqrt{Rr}}\zeta'_J(R,r,\gamma) \\
    \int R dR\, r dr\, d\gamma  \zeta_J^*(R,r,\gamma) &\left[\frac{1}{R}\frac{\partial}{\partial R}R\frac{\partial}{\partial R} + \frac{1}{r}\frac{\partial}{\partial r}r\frac{\partial}{\partial r} \right]\zeta_J(R,r,\gamma)  \nonumber \\
    &= \int dRdrd\gamma 
    \zeta^{'*}_J(R,r,\gamma) \left[\frac{\partial^2}{\partial R^2} +\frac{1}{4R^2}+ \frac{\partial^2}{\partial r^2} + \frac{1}{4r^2}\right]  \zeta'_J(R,r,\gamma) 
\end{align}

In 3D, a similar rescaling is performed via,\cite{Schatz1976_3D}
\begin{align}
    \xi^J_{\ell\Omega}(R,r) & = \frac{1}{Rr}\xi'^J_{\ell\Omega}(R,r) \\
    \sum_{\ell\Omega}\int R^2dR\, r^2dr\, &\xi^{J*}_{\ell\Omega}(R,r) \Bigg[\frac{1}{R^2}\frac{\partial}{\partial R}R^2\frac{\partial}{\partial R} +\frac{1}{r^2}\frac{\partial}{\partial r}r^2 \frac{\partial}{\partial r} \Bigg]\xi^J_{\ell\Omega}(R,r)   \nonumber \\
    &= \sum_{\ell\Omega}\int dR dr \xi'^{J*}_{\ell\Omega}(R,r) \left[\frac{\partial^2}{\partial R^2} +\frac{\partial^2}{\partial r^2} \right]  \xi'^J_{\ell\Omega}(R,r)
\end{align}
    
In the 3D, calculation, $N_\ell = 50$ as the spherical harmonic grid is much more efficient than a physical grid. For the calculating the potential in the spherical harmonic basis, as defined in Eq. \ref{eq:Vsph}, we choose $N_\gamma=4000$ for full convergence. The extents for $R$ were set using the same interpolation from Table \ref{tab:extents} as in the exact calculation.

\paragraph{For the PS code:} The `electron' like kinetic energy is implemented on cartesian grids $\hat{x}\hat{y}\hat{z}$ with an 11th-order finite difference based kinetic energy. A grid size of $N_R = 91 \ \text{and}\   N_x= N_y = N_z = 90$ such that a BO calculation using the cartesian quadrature numerically agrees with a BO calculation using the polar grid quadrature in 2D or spherical harmonic in 3D of the exact calculation.

\begin{table}[]
    \centering
    \begin{tabular}{|c|c|c|}
    \hline
         $\mu_R$ (amu)& $R_{min}$ (au) & $R_{max}$ (au)  \\
         \hline
         2 & 3.5 & 6.4 \\
         10 & 3.9 & 5.9 \\
         50 & 4.0 & 5.5 \\
         100 & 4.1 & 5.4 \\
         1000 & 4.2 & 5.2 \\
         \hline
    \end{tabular}
    \caption{Reference boundary extents for the $R$ coordinate. Extents for intermediate masses were interpolated from this set.}
    \label{tab:extents}
\end{table}
\section{Review of Ref. \cite{Tao2025_basis_free}:  General form for $\Gamma$ \label{appendix:gamma''}}

For the sake of completeness, we list here the proposed form for $\hbG$ for an arbitrary set of $N_A$ atoms, where many rotational and torsional modes exist (for more details see Ref. \cite{Tao2025_basis_free}). To satisfy the constraints in Fig. \ref{fig:dgammasym}, we set:
\begin{align}
     \hat{\bm \Gamma}'_A = \frac{-i}{2\hbar}\left\{\hat{\Theta}_A, \hat{\uvp}_e\right\}, 
\end{align}
which is unchanged from Eq. \ref{eq:etf_define} in the main text. We generalize $\hbG''$ as,
\begin{align}
        \hat{\bm \Gamma}_A'' &= -\sum_B\zeta_{AB}\left(\bm{X}_A -\bm{X}^0_{B}\right)\times \left(\bm{K}_B^{-1}\hat{\bm J}_B\right)\label{eq:erf_final}.
\end{align}
Here, we have introduced the factors $\zeta_{AB}$, $\bm X^0$, $\bm K_B$ to induce locality, such that a torsion in one part of the molecule, is not felt by electrons far away in another part of another molecule.
\begin{align}
    \hat{\bm J}_B &= \frac{1}{2i\hbar}\left((\hat{\bm r}-\bm{X}_B)\times \hat \Theta_B(\hat{\bm r} )\hat{\bm p} +(\hat{\bm r} -\bm{X}_B)\times\hat{\bm p} \hat \Theta_B(\hat{\bm r})\right), \\
    \bm{X}_{B}^0 &= \frac{\sum_A\zeta_{AB}\bm{X}_A}{\sum_A\zeta_{AB}},\\
    \label{eq:KB}
    \bm{K}_B &= \sum_A\zeta_{AB}\left((\bm{X}_A^\top\bm{X}_A-\bm{X}_B^{0\top}\bm{X}_B^0)\mathcal{I}_3-\bm{X}_A\bm{X}_A^\top+\bm{X}_B^0\bm{X}_B^{0\top}\right) \\
    \zeta_{AB} &=M_A e^{-|\bm{X}_A-\bm{X}_B|^2/\beta_{AB}^2}.
\end{align}
While the definition of $\hat{\bm J}_B$ is unchanged from the definitions in Eq. \ref{eq:erf_J1_define} and \ref{eq:erf_J2_define}, $\bm X^0_B$ is defined as the local center of mass for a rotation around atom B, and $\bm K_B$ is taken to be the local moment of inertia around atom B (simplifying to the true diatomic moment of inertia in Eq. \ref{eq:mom_inertia} when $\zeta_{AB} = M_A$ and all $\bm X^0 = 0$), an essential property in determining how electrons near atom B respond to a nuclear rotation nearby. \nadine{We note that there are occasions where $\bm K_B$ becomes singular (in diatomics or when three or more atoms become colinear) and cannot be inverted. In such cases, one has to take care to invert only the non-singular submatrix of $\bm K_B$ as detailed in Ref. \cite{Qiu2024_erf}.} The locality function $\zeta$ sets length scale over which we expect local torsions and rotations to affect the electronic density (to be determined for each system); here, this length scale is set by the factor $\beta_{AB}$.

In the presence of spin, there can be an additional term,  $\hbG'''$, that rotates the electronic spin into the molecular body frame, such that $\bP\cdot \hbG'''$ creates a spin-Coriolis term in the Hamiltonian\cite{Bradbury2025_SpinCoriolis}
\begin{eqnarray}
    \hat{\bm \Gamma}_A''' &=& \sum_B\zeta^{(s)}_{AB}\left(\bm{X}_A -\bm{X}^0_{B}\right)\times \left(\bm{K}_B^{-1}\hat{\bm J}_B^{(s)}\right),\label{eq:erf_final_spin}\\
    \hat{\bm J}_B^{(s)} &=& \frac{-1}{i\hbar}\left(\hat{\bm s} \hat{\Theta}_B(\hat{\bm r})\right).
\end{eqnarray}
Here, $\zeta^{(s)}_{AB}$ sets the correlation length for spin, and one need not require that the latter function equal $\zeta_{AB}$. $\hbG'''$ was not used in the main text.

\section{Final equations and results from 2-dimensional calculations\label{appendix:2D_data}}
While all of the data above has been presented for the 3D case, similar data can be extracted (and far more easily) for the 2D case following the same general analysis as above. Indeed, one can run a 2D simulation also using our publicly available GPU package. 
The phase space Hamiltonian in 2 dimensions (with coordinates defined in Fig. \ref{fig:bfc_diagram}) is:
\begin{eqnarray}
\hat H_{\PS}  &=& \frac {\Big(P_x- i\hbar \hat \Gamma_x\Big)^2} {2\mu_{R}}+\frac {\Big( P_y - i\hbar \hat \Gamma_y\Big)^2} {2\mu_R}+ \hat{H}_{\el}(\hat{x},\hat{y}) \\
    &=& \frac {\Big(- P_\vartheta /R \sin\vartheta_R+ P_R\cos\vartheta_R- i\hbar \hat \Gamma_x\Big)^2} {2\mu_{R}}+\frac {\Big(  P_\vartheta/R \cos\vartheta_R+  P_R \sin\vartheta_R - i\hbar \hat \Gamma_y\Big)^2} {2\mu_{R}}+ \hat H_{\el}(\hat x,\hat y)\\
&=& \frac{1}{2\mu_{R}}\Big(\frac{P_\vartheta^2}{R^2}+P_R^2\Big)
    -\frac{i\hbar}{R\mu_{R}}(- P_\vartheta \sin\vartheta_R+R P_R\cos\vartheta_R)\hat{\Gamma}_x-\frac{i\hbar}{R\mu_{R}}(  P_\vartheta \cos\vartheta_R+ R P_R\sin\vartheta_R)\hat{\Gamma}_y \nonumber \\&& -\frac{\hbar^2}{2\mu_{R}}\left(\hat{\Gamma}_x^2 + \hat{\Gamma}_y^2 \right) + \hat H_{\el}(\hat x,\hat y)
\end{eqnarray}
For a choice of gauge with $\vartheta_R = 0$:
\begin{eqnarray}
\hat H_{\PS}  
&=& \frac{1}{2\mu_{R}}\Big(\frac{P_\vartheta^2}{R^2}+P_R^2\Big)
    -\frac{i\hbar}{\mu_R}P_R\hat{\Gamma}_x
    -\frac{i\hbar}{\mu_{R}}  \frac{P_\vartheta}{R} \hat{\Gamma}_y  -\frac{\hbar^2}{2\mu_{R}}\left(\hat{\Gamma}_x^2 + \hat{\Gamma}_y^2 \right) + \hat H_{\el}(\hat x,\hat y)
\end{eqnarray}

Lastly, as discussed in Section \ref{sec:PS_Ptheta}, we can take $P_\vartheta \to \hbar J$ and write the Hamiltonian for rotational state $J$ (analogous to Eq \ref{eq:compare:bunker}) as:
\begin{align}
    \hat H_{\PS}^J  
&= \frac{1}{2\mu_{R}}\Big(\frac{\hbar^2J^2}{R^2}+P_R^2\Big)
    -\frac{i\hbar}{\mu_R}P_R\hat{\Gamma}_x
    -\frac{i\hbar^2}{\mu_{R}}  \frac{J}{R} \hat{\Gamma}_y  -\frac{\hbar^2}{2\mu_{R}}\left(\hat{\Gamma}_x^2 + \hat{\Gamma}_y^2 \right) + \hat H_{\el}(\hat x,\hat y).
\end{align}
The electron translations  factors $\hat{\Gamma}'_x,\hat{\Gamma}'_y$ are defined as:
\begin{eqnarray}
    \hat{\Gamma}'_x &=&
     \frac{1}{2i\hbar}
    \Big\{\hat{\Theta}_R,\hat{ {\up}}_x\Big\}  \\
    \hat{\Gamma}'_y &=& 
     \frac{1}{2i\hbar}
    \Big\{\hat{\Theta}_R,\hat{ {\up}}_y\Big\} 
\end{eqnarray}
The electronic angular momentum for the two-dimensional rotating molecule points out of plane, the electron rotation factor projects it back into the plane, thus $\hbG''$ is 
\begin{eqnarray}
    \hat{\Gamma}''_y &=& \frac{1}{Ri\hbar }\left(\hat{r}_x\hat{p}_y-\hat{r}_y\hat{p}_x -\frac{R}{2 }\Big\{\hat{\Theta}_R, \hat{p}_y\Big
\}\right)
\end{eqnarray}
where $\hat{\Theta}_R$ is defined in Eq. \ref{eq:thetaR}.
Thus, in the end, we find that the total $\hG_x$ and $\hG_y$ are of the form:
\begin{eqnarray}
\hat{\Gamma}_x &=&
     \frac{1}{2i\hbar}
    \Big\{\hat{\Theta}_R,\hat{ {\up}}_x\Big\}  \\
\hat{\Gamma}_y &=& 
     \frac{1}{2i\hbar}
    \Big\{\hat{\Theta}_R,\hat{ {\up}}_y\Big\} + \frac{1}{Ri\hbar }\left(\hat{r}_x\hat{p}_y-\hat{r}_y\hat{p}_x -\frac{R}{2 }\Big\{\hat{\Theta}_R, \hat{p}_y\Big\} \right) = \frac{\hat l_z}{i\hbar R}   
\end{eqnarray}

\begin{figure*}
    \centering
    \includegraphics[]{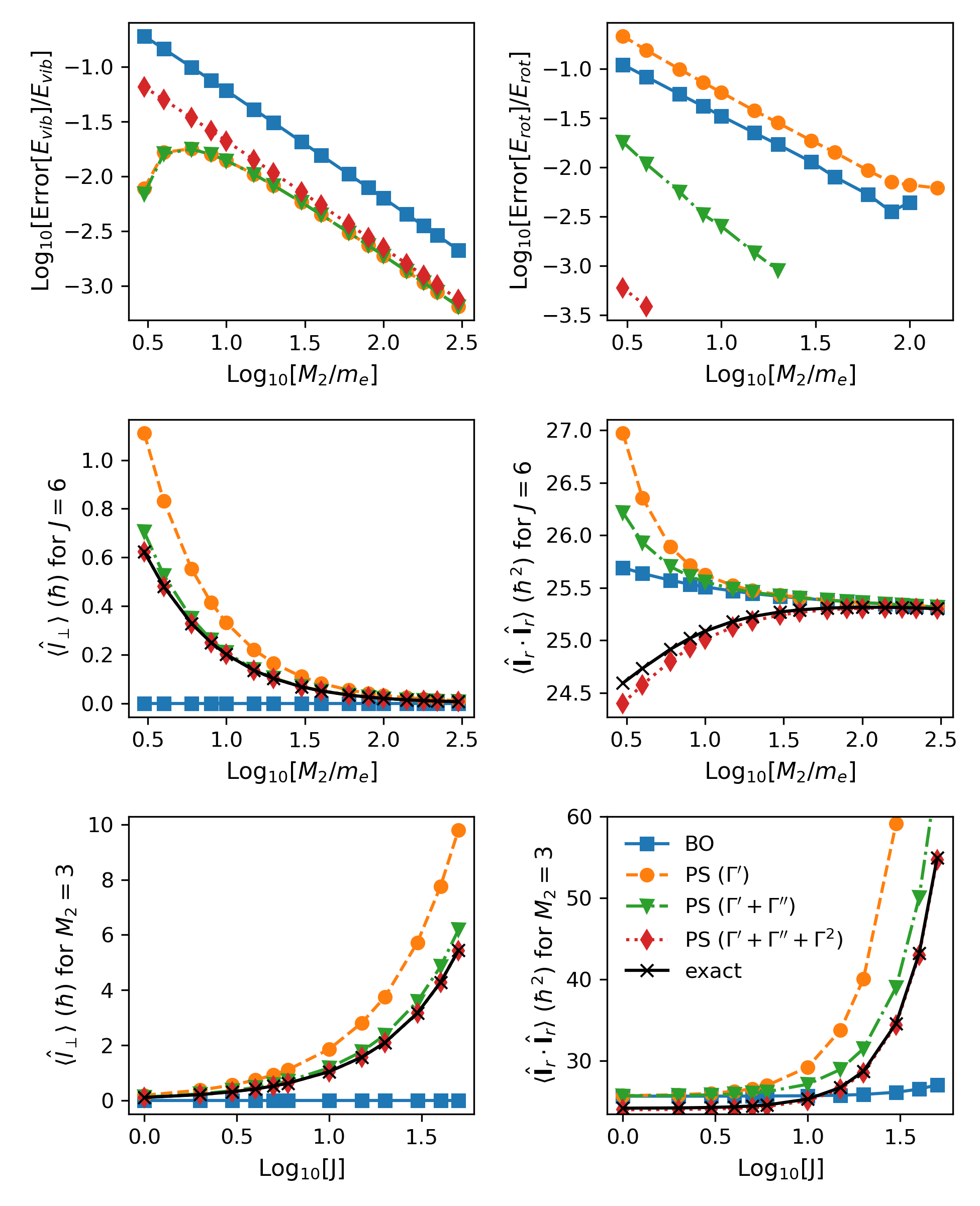}
    \caption{Representative data from the 2D calculations. In the top row we plot error in the vibrational state gap (top-left), and the rotational transition $J=6\to0$ (top-right) as a function of the mass. This log-log plot is analogous to the data in Fig. \ref{fig:energy_gaps}. In the middle row, we plot the angular momentum (perpendicular (middle-left) and square (middle-right) at fixed $J=6$ vs the log of the mass ratio, analogous to the data in Fig. \ref{fig:ang_mass} On the bottom row, we plot this same data (now at fixed $M_2=3$) vs the $\mathrm{log}_{10}[J]$, analogous to the data in Fig \ref{fig:ang_J}. }
    \label{fig:2D_fig}
\end{figure*}

In Fig. \ref{fig:2D_fig}, we plot  representative data for the 2D-calculation. We observe trends in the errors of the vibrational and rotational gaps that are very similar to the 3D data shown in Fig. \ref{fig:energy_gaps}. The only notable exception is that the slope of the error of the rotational gaps for PS theories that include $\mathbf{\hat{\Gamma}}''$ appears to be more similar to the slope seen for BO theory than in 3D. We also find that the $\mathbf{\hat{\Gamma}}^2$ term does not provide an additional improvement in the rotational energy error in comparison to 3D. When it comes to the angular momenta, we find that PS performs similarly well in 2D and in 3D. We note that, due to the fact that $J$ is just a parameter in the 2D calculation (leading the lack of the $\Omega$ dimension in the wavefunction shape), we can access much higher $J$ values in 2D vs 3D. Here we can demonstrate that even on a $\mathrm{log}[J]$ scale, PS performs exceptionally well for the amount of angular momentum in the electronic ground state.

Note that, in all cases, PS electronic structure is quite accurate and far surpasses what is possible with BO theory.

\section{Electron Rotation Factor Calculation in 3-dimensions (Derivation of Eqs. \ref{eq:gy_derive}-\ref{eq:gz_derive}) }\label{appendix:erf_cals}
Using Eqs. \ref{eq:erf_define}- \ref{eq:erf_J2_define} we can write $\hbG_1'',\hbG_2''$ as
\begin{eqnarray}
    \hbm{\Gamma}_1'' &=&-M_1\bm{X}_1\times \left(\bm{I}^{-1}\hbm{J}_1\right)-M_1\bm{X}_1\times \left(\bm{I}^{-1}\hbm{J}_2\right) \label{eq:Gamma''1}\\
    \hbm{\Gamma}_2'' &=& -M_2\bm{X}_2 \times \left(\bm{I}^{-1}\hbm{J}_1\right)-M_2\bm{X}_2\times \left(\bm{I}^{-1}\hbm{J}_2\right)\label{eq:Gamma''2}
\end{eqnarray}
    Since we choose internuclear axis to align with the x-axis in body frame, the nuclear center of mass coordinates Eqs. \ref{eq:R_NCM}-\ref{eq:R12} can be written in component form as:
\begin{align}
    \uvX_1 = \begin{pmatrix}
    X_{1x} \\
    X_{1y}\\
    X_{1z}
\end{pmatrix} =  R\frac{\mu_R}{M_1}\begin{pmatrix}
    1 \\
    0 \\
    0
\end{pmatrix} && \uvX_2 = \begin{pmatrix}
    X_{2x} \\
    X_{2y}\\
    X_{2z}
\end{pmatrix} =-R\frac{\mu_R}{M_2}\begin{pmatrix}
    1 \\
    0 \\
    0
\end{pmatrix}
\end{align}
Next we define angular momentum with respect to the nuclear coordinate as follows:
\begin{align}
    \hbm{J}_1 &= \frac{1}{2i\hbar}\left((\hbm{r}-\bm{X}_1)\times(\hat{\Theta}_1\hbm{p}+\hbm{p}\hat{\Theta}_1)\right)\\
     &= \frac{1}{2i\hbar}\begin{pmatrix}
  \hat{r}_y (\hat{\Theta}_1\hat{p}_z+\hat{p}_z\hat{\Theta}_1)-\hat{r}_z(\hat{\Theta}_1\hat{p}_y+\hat{p}_y\hat{\Theta}_1)\\
  \\
  \hat{r}_z (\hat{\Theta}_1\hat{p}_x+\hat{p}_x\hat{\Theta}_1)-\Big(\hat{r}_x -R\frac{\mu_R}{M_1}\Big)(\hat{\Theta}_1\hat{p}_z+\hat{p}_z\hat{\Theta}_1)\\
  \\
  \Big(\hat{r}_x-R\frac{\mu_R}{M_1} \Big)(\hat{\Theta}_1\hat{p}_y+\hat{p}_y\hat{\Theta}_1)-\hat{r}_y(\hat{\Theta}_1\hat{p}_x+\hat{p}_x\hat{\Theta}_1)
\end{pmatrix}\\
    \hbm{J}_2 &= \frac{1}{2i\hbar}\left((\hbm{r}-\bm{X}_2)\times(\hat{\Theta}_2\hbm{p}+\hbm{p}\hat{\Theta}_2)\right)\\
    &=\frac{1}{2i\hbar}\begin{pmatrix}
  \hat{r}_y (\hat{\Theta}_2\hat{p}_z+\hat{p}_z\hat{\Theta}_2)-\hat{r}_z(\hat{\Theta}_2\hat{p}_y+\hat{p}_y\hat{\Theta}_2)\\
  \\
  \hat{r}_z (\hat{\Theta}_2\hat{p}_x+\hat{p}_x\hat{\Theta}_2)-\Big(\hat{r}_x +R\frac{\mu_R}{M_2}\Big)(\hat{\Theta}_2\hat{p}_z+\hat{p}_z\hat{\Theta}_1)\\
  \\
  \Big(\hat{r}_x+R\frac{\mu_R}{M_2} \Big)(\hat{\Theta}_2\hat{p}_y+\hat{p}_y\hat{\Theta}_1)-\hat{r}_y(\hat{\Theta}_2\hat{p}_x+\hat{p}_x\hat{\Theta}_1)
\end{pmatrix}
\end{align}
Further, the moment of inertia is defined in Eq. \ref{eq:Iinv} which gives us the final expressions for electron rotation factors in Eqs.\ref{eq:Gamma''1}-\ref{eq:Gamma''2}:

\begin{align}
    \hat{\bm{\Gamma}}_1^{''} &= -M_1 R\frac{\mu_R}{M_1} \begin{pmatrix}
    1\\
    0 \\
    0
\end{pmatrix}\times \left(\frac{1}{R^2\mu_R}
\begin{pmatrix}
    0 & 0 & 0 \\
    0 & 1 & 0\\
    0 & 0 & 1
\end{pmatrix} \cdot \begin{pmatrix}
    \hat{J}_{1x}+\hat{J}_{2x} \\
    \hat{J}_{1y}+\hat{J}_{2y}\\
    \hat{J}_{1z}+\hat{J}_{2z}
\end{pmatrix}\right)
\\
&=-\frac{1}{R}\begin{pmatrix}
    0 \\
    -\hat{J}_{1z}-\hat{J}_{2z} \\
    \hat{J}_{1y}+\hat{J}_{2y}
\end{pmatrix} = \frac{1}{Ri\hbar}\begin{pmatrix}
    0 \\
    \hat{r}_x\hat{p}_y-\hat{r}_y\hat{p}_x-\frac{R}{2} \Big\{\hat{\Theta}_R,\hat{p}_y\Big\}\\
     -(\hat{r}_z\hat{p}_x-\hat{r}_x\hat{p}_z) -\frac{R}{2} \Big\{\hat{\Theta}_R, \hat{p}_z\Big\}
\end{pmatrix}
\\
    \hat{\bm{\Gamma}}_2^{''} &= -\hat{\bm{\Gamma}}_1^{''}
\end{align}
where $\hat{\Theta}_R$ is defined in Eq. \ref{eq:thetaR}. Finally $\hat{\bm \Gamma}_R$ in Eq.\ref{eq:gamma_rel} can be written as: 
\begin{align}
     \hat{\bm{\Gamma}}_R = \frac{M_2\hat{\bm{\Gamma}}_1-M_1\hat{\bm{\Gamma}}_2}{M_2+M_1}
\end{align}
Eqs. \ref{eq:gy_derive}-\ref{eq:gz_derive} now follow.

\section{On the Definition of the Nonabelian Curl for Electronic Operators\label{sec:appendix_curl}
}
The definition of the nonabelian curl in Eq. \ref{eq:def_jodd} might not be familiar to the reader. After all, the nonabelian curl is usually defined with respect to a set of adiabatic states, rather than to an arbitrary vector field that operates on the entire electronic hilbert space. To that end, consider a semiclassical system (with classical nuclei and quantum electrons), with a phase space electronic Hamiltonian of the form in Eq. \ref{eq:gamma_general}, which we repeat here:
\begin{align}
   \hH_{PS} = \frac{\bP^2}{2M} + \hH_{el} -i\hbar \hbG \cdot \frac{\bP}{M}  -\hbar^2 \frac{\hbG \cdot \hbG}{2M}
   \label{eq:again}
\end{align}
At this point, $\hbG$ can be an arbitrary function of nuclear position, $\hbG(\bR)$. We will assume that the Hamiltonian in Eq. \ref{eq:again} is defined in a total subspace that is invariant to nuclear position and momentum.

We will now develop the relevant \joe{semiclassical} equations of motion for this system. Let the electronic density matrix be $\hat{\rho}(R,P)\equiv \hat{\rho}$.
The  total energy is:
\begin{align}
 E &= \joe { \int dR \int dP } \; \mbox{tr} \left( \hat{\rho} \hH_{PS} \right) =  \joe {\int dR \int dP } \; \frac{\bP^2}{2M} + \mbox{tr}\left( \hat{\rho} \left( \hH_{el} -i\hbar \hbG \cdot \bP/M  -\hbar^2 \hbG \cdot \hbG/(2M)\right)\right) 
\end{align}
For such a system with quantum electrons and classical nuclei, we further stipulate that all equations of motion will be in the Heisenberg representation and of the form:
\begin{align}
     \frac{d\hat{O}}{dt} &= \frac{i}{\hbar}\left[ \hH_{PS},\hat{O}\right] -\frac{1}{2} \left(  \left\{\hH_{PS},\hat{O} \right\} - \left\{\hat{O},\hH_{PS} \right\} \right) \label{eq:semiclassical_heisenberg_eom}
\end{align}

A proof that one can work equivalently in the Heisenberg and Schrodinger representations within a mixed quantum classical framework is presented below. \joe {The framework presented so far is equivalent to the quantum-classical Liouville equation\cite{kapral:1999:jcp,martens:1997:partwig}. }

\joe{Now, we can also take one step further and put the above equations of motion in Ehrenfest form by working locally at every $R$ and $P$ point in phase space. Thus, we can approximate the relevant equations of motion to be:}
\begin{align}    
     \dot{R}_{I  \alpha} &= \mbox{tr} \left( \hat{\rho}\frac{d}{dt} \left( R_{I \alpha} \otimes \hat{I}_e \right) \right) = \frac{\partial E}{\partial P_{I \alpha}} \\
    \dot{P}_{I \alpha} &= \mbox{tr} \left(\hat{\rho} \frac{d}{dt} \left( P_{I \alpha} \otimes \hat{I}_e \right) \right) = -\frac{\partial E}{\partial R_{I \alpha}} \\
\frac{d\hat{\Gamma}^{I \alpha}}{dt} &= \frac{i}{\hbar}\left[ \hH_{PS},\hat{\Gamma}^{I \alpha}\right] -\frac{1}{2} \left(  \left\{\hH_{PS},\hat{\Gamma}^{I \alpha} \right\} - \left\{\hat{\Gamma}^{I \alpha},\hH_{PS} \right\} \right) \\
 & = \frac{i}{\hbar}\left[ \hH_{PS},\hat{\Gamma}^{I \alpha}\right] +\sum_{J \beta} \frac{1}{2M_J} \left( \left( P_{J \beta} - i\hbar \hat{\Gamma}^{J \beta} \right) \frac{\partial \hat{\Gamma}^{I \alpha}}{\partial R_{J \beta}}  +   \frac{\partial \hat{\Gamma}^{I \alpha}}{\partial R_{J \beta}} \left( P_{J \beta} - i \hbar \hat{\Gamma}^{J \beta} \right) \right)
\end{align}
It then follows that:
\begin{align}
\label{eq:kinetic_momentum}
     \dot{R}_{I \alpha} &=  \frac{P_{I \alpha}}{M_I} -\frac{i\hbar}{M_I} \mbox{tr}\left( \hat{\rho} \hat{\Gamma}^{I \alpha} \right)   \\
    \dot{P}_{I \alpha} &= - \mbox{tr}\left( \hat{\rho}  \left( \frac{\partial \hH_{el}}{\partial R_{I \alpha}}  -i\hbar \frac{\partial \hbG}{\partial R_{I \alpha}} \cdot \frac{\bP}{M_I}  - \frac{\hbar^2}{2M_I} \left( \frac{\partial \hbG}{\partial R_{I \alpha}} \cdot \hbG
    +
     \hbG \cdot
    \frac{\partial \hbG}{\partial R_{I \alpha}} 
    \right) \right)  \right) 
\end{align}
Moreover, if we take the second derivative of the velocity, we find:
\begin{align}
     \ddot{R}_{I \alpha} &=  \frac{\dot{P}_{I \alpha}}{M_I} -\frac{i\hbar}{M_I} \mbox{tr}\left( \hat{\rho} \frac{d}{dt} \hat{\Gamma}^{I \alpha} \right) \\
     &= -\frac{1}{M_{I \alpha}} \mbox{tr}\left( \hat{\rho}  \left( \frac{\partial \hH_{el}}{\partial R_{I \alpha}}  -i\hbar \sum_{J \beta} \frac{\partial \hat{\Gamma}^{J \beta}}{\partial R_{I \alpha}} \frac{P_{J \beta}}{M_{J}}  -\hbar^2 \sum_{J \beta} \frac{1}{2M_J} \left( \frac{\partial  \hat{\Gamma}^{J \beta}}{\partial R_{I \alpha}} 
     \hat{\Gamma}^{J \beta} + \hat{\Gamma}^{J \beta}  
     \frac{\partial  \hat{\Gamma}^{J \beta}}{\partial R_{I \alpha}}
     \right)  \right)  \right)  \\
      \nonumber 
 &  \nadine{+} \frac{1}{M_I} \mbox{tr} \left( \hat{\rho} \left[ \hH_{PS},\hat{\Gamma}^{I \alpha}\right] \right) - \sum_{J \beta} \frac{i\hbar}{2M_IM_J} \mbox{tr} \left ( \hat{\rho} \left( \left( P_{J \beta} -\nadine{i\hbar} \hat{\Gamma}^{J \beta} \right) \frac{\partial \hat{\Gamma}^{I \alpha}}{\partial R_{J \beta}}  +   \frac{\partial \hat{\Gamma}^{I \alpha}}{\partial R_{J \beta}} \left( P_{J \beta} - \nadine{i\hbar}\hat{\Gamma}^{J \beta} \right) \right) \right)
\\
      &= -\frac{1}{M_{I}} \mbox{tr}\left( \hat{\rho}  \left( \frac{\partial \hH_{el}}{\partial R_{I \alpha}}  -i\hbar \sum_{J \beta} \frac{\partial \hat{\Gamma}^{J \beta}}{\partial R_{I \alpha}} \frac{P_{J \beta}}{M_{J}}  -
      \hbar^2 \sum_{J \beta} \frac{1}{2M_J}
      \left( \frac{\partial  \hat{\Gamma}^{J \beta}}{\partial R_{I \alpha}} 
     \hat{\Gamma}^{J \beta} + \hat{\Gamma}^{J \beta}  
     \frac{\partial  \hat{\Gamma}^{J \beta}}{\partial R_{I \alpha}}
     \right)   \right) \right)     \\
     \nonumber
     & +\frac{1}{M_I} \mbox{tr}\left(  \hat{\rho} \left[\hH_{el},\hat{\Gamma}^{I \alpha}\right]  \right)
      - \sum_{J \beta} \frac{i\hbar}{M_I M_J} \mbox{tr}\left( \hat{\rho} \left[\hat{\Gamma}^{J \beta} ,\hat{\Gamma}^{I \alpha}\right]  \right) P_{J \beta}
      \\ 
      \nonumber
      &
      -\sum_{J \beta} \frac{\hbar^2}{2M_I M_J} \mbox{tr}\left( \hat{\rho} \left[ 
      \hat{\Gamma}^{J \beta} 
     \hat{\Gamma}^{J \beta}
      ,\hat{\Gamma}^{I \alpha} \right] \right) 
      \\
      &
      \nonumber
- \sum_{J \beta} \frac{i \hbar}{2M_IM_J} \mbox{tr} \left ( \hat{\rho} \left( \left( P_{J \beta} - i \hbar \hat{\Gamma}^{J \beta} \right) \frac{\partial \hat{\Gamma}^{I \alpha}}{\partial R_{J \beta}}  +   \frac{\partial \hat{\Gamma}^{I \alpha}}{\partial R_{J \beta}} \left( P_{J \beta} - i \hbar\hat{\Gamma}^{J \beta} \right) \right) \right)
\end{align}
If we simplify the equations above, we find:
\begin{align}
M_I \ddot{R}_{I \alpha} &=
F_{I \alpha}(\bR) - \frac{i \hbar}{2} 
\sum_{J \beta} \frac{1}{M_J} \mbox{tr} \left( \hat{\rho}  \left( \hat{\Omega}_{J \beta, I \alpha}  \left(  P_{J \beta}  - i \hbar \hat{\Gamma}^{J \beta}  \right) + \left(  P_{J \beta}  - i \hbar \hat{\Gamma}^{J \beta} \right) \hat{\Omega}_{J \beta, I \alpha}    \right) \right)\\
& = F_{I \alpha}(\bR) - \frac{i \hbar}{2} 
\sum_{J \beta} \mbox{tr} \left( \hat{\rho}  \left( \hat{\Omega}_{J \beta, I \alpha}  \left(  \frac{\partial \hH_{PS}}{\partial P_{J \beta}}  \right) + \left(  \frac{\partial \hH_{PS}}{\partial P_{J \beta}}  \right)   \hat{\Omega}_{J \beta, I \alpha}  \right) \right)
\\
\Omega_{J \beta, I \alpha} &=
\left(
\frac{\partial \hat{\Gamma}^{I \alpha}}{\partial R_{J \beta}}
-
\frac{\partial \hat{\Gamma}^{J \beta}}{\partial R_{I \alpha}}
- \left[ \hat{\Gamma}^{I \alpha},\hat{\Gamma}^{J \beta} \right]
\right)  \\
F_{I \alpha}(\bR) 
      & = - \mbox{tr}\left( \hat{\rho}  \left( \frac{\partial \hH_{el}}{\partial R_{I \alpha}}
      - \left[\hH_{el}, \hat{\Gamma}^{I \alpha}
\right]  \right) \right) 
\end{align}
Thus, \joe{with Ehrenfest dynamics}, we find the same form for the curvature (i.e. magnetic field) as we conjectured  in Eq. \ref{eq:def_jodd} and  in agreement with the usual nonabelian Berry  curvature expression (but now reinterpreted).

At this juncture,  note that the manipulations above are quite general and one could have chosen to work in an adiabatic basis, take $\hH_{el}$ to be diagonal, and set  $\hat{\Gamma}$ to be the derivative couplings. Such a scenario would correspond to performing an adiabatic transformation before taking the Wigner transform\cite{izmaylov:2014:qcle_berry} and taking a mean-field approximation.   In such a case,  in a complete basis, $\hat{\Omega}$ would vanish and we would recover the usual Ehrenfest equations of motion.  In an incomplete basis, however, we would find that the equations of motion have become more complicated and now involve pseudomagnetic forces.  See also Refs. \cite{takatsuka:2005:jcp,takatsuka:2011:pccp_review_nonadiabatic,krishna:2007:ehr_plus_berry,coraline:2024:jcp:ehrenfest_conserve} for a more complete description.  For either a complete or incomplete basis, the functions $\bP_{tot} = \sum_I M_I \dot{\bR}_I + \mbox{tr} \left( \hat{\rho}\, \hbp \right)$ and $\bL_{tot} = \sum_I M_I \dot\bR_I \times \bR_I + \mbox{tr} \left( \hat{\rho}\, \hbl \right)$
are conserved, corresponding to total linear and angular momentum conservation.

For our purposes, however, we can take $\hH_{el}$ in Eq. \ref{eq:H_el} to be the standard electronic Hamiltonian and $\hbG$ to be the operator defined by Eqs. \ref{eq:etf_define} and \ref{eq:erf_define}. In this case, because the nonabelian curl of $\hbG$ is nonzero, we will not recapitulate standard Ehrenfest dynamics if we work in a complete basis. That being said, the data presented in this manuscript suggests that the Hamiltonian in Eq. \ref{eq:again}  is quite meaningful and that, if work in a truncated basis of the lowest few levels, we can outperform more conventional results obtained with the standard electronic Hamiltonian. Moreover,   the physical meaning of the magnetic force in $\Omega$ is always the same, highlighting the fact that phase space electronic structure theory does indeed recover  the importance aspects of the MSW curl.

\bigskip

\subsection{Proof of the Equivalence of Heisenberg and Schrodinger representations for a mixed quantum-classical Hamiltonian}
The  only remaining technical hurdle is to demonstrate that the Schrodinger and Heisenberg pictures -- which are well known and equivalent in exact quantum mechanics -- can also be applied and are still equivalent for semiclassical mechanics and equations of motion of the form Eq. \ref{eq:semiclassical_heisenberg_eom}. 
To that end, let us now show that if we work with any operator $A$, we will find that we achieve the same trace whether we propagate $A$ or $\hat{\rho}$ in time. 
To prove this statement, note that for any operators $A$ and $B$,
\begin{align}
\mbox{tr}\left(A[B,C]\right) &=  \mbox{tr}\left([A,B]C \right) = - \mbox{tr}\left(C [B,A] \right) \\
\int dR \int dP \; \mbox{tr} \left( \hat{A}  \left(  \left\{\hat{B},\hat{C} \right\}  \right) \right) &= 
\int dR \int dP \; \mbox{tr} \left( \hat{A}  \left(  \frac{\partial \hat{B}}{\partial R} \frac{\partial \hat{C}}{\partial P}  
-  \frac{\partial \hat{B}}{\partial P} \frac{\partial \hat{C}}{\partial R}  
  \right) \right) 
\\
&= - \int dR \int dP \; \mbox{tr} \left(  \hat{C}   \frac{\partial}{\partial P} \left( \hat{A}  \frac{\partial \hat{B}}{\partial R} \right) 
-   \hat{C} \frac{\partial}{\partial R}  \left(  \hat{A}  \frac{\partial \hat{B}}{\partial P} \right) 
 \right)  \\
&= - \int dR \int dP \; \mbox{tr} \left(  \hat{C}   \frac{\partial \hat{A}}{\partial P}   \frac{\partial \hat{B}}{\partial R}
-   \hat{C} \frac{\partial \hat{A} }{\partial R}    \frac{\partial \hat{B}}{\partial P} 
 \right)  \\
&= 
\int dR \int dP \; \mbox{tr} \left( \hat{C}  \left(  \left\{\hat{A},\hat{B} \right\}  \right) \right)      
\end{align}
Above we have assumed that the integral is bounded so we can integrate by parts and the surface term is zero. A slightly different integration by parts also shows that:

\begin{align}
\int dR \int dP \; \mbox{tr} \left( \hat{A}  \left(  \left\{\hat{B},\hat{C} \right\}  \right) \right)
&= 
 \int dR \int dP \; \mbox{tr} \left( \hat{B}  \left(  \left\{\hat{C},\hat{A} \right\}  \right) \right)   
\end{align}
With these results, we can compute:
\begin{align}
\frac{d}{dt}  \int dR \int dP \; \mbox{tr} \left( \hat{\rho} A(t) \right) &=  \int dR \int dP \; \mbox{tr} \left( \hat{\rho} \left(
 \frac{i}{\hbar}\left[ \hH_{PS},\hat{A}\right]-\frac{1}{2} \left(  \left\{\hH_{PS},\hat{A} \right\} - \left\{\hat{A},\hH_{PS} \right\} \right) \right)\right)
 \\
&= \int dR \int dP \; \mbox{tr} \left( \hat{A} \left(
 \frac{-i}{\hbar}\left[ \hH_{PS},\hat{\rho}\right] + \frac{1}{2}    \left( \left\{ \hH_{PS}, \hat{\rho}  \right\} - \left\{\hat{\rho}, \hH_{PS} \right\} \right) \right) \right)
\end{align}
Thus, propagating an observable $\hat{A}$ in the Heisenberg framework while keeping $\hat{\rho}$ fixed remains equivalent to a Schrodinger framework for which $\hat{A}$ is kept fixed and we propagate $\hat{\rho}$ with a minus sign, 
recovering the standard quantum classical Liouville equation\cite{kapral:1999:jcp,martens:1997:partwig}:

\begin{align}
     \frac{d\hat{\rho}}{dt} &= \frac{-i}{\hbar}\left[ \hH_{PS},\hat{\rho}\right] 
     + \frac{1}{2} \left(  \left\{\hH_{PS},\hat{\rho} \right\} - \left\{\hat{\rho},\hH_{PS} \right\} \right)
\end{align}

\vspace{\columnsep}
\twocolumngrid


%

\end{document}